\documentclass[10pt,journal,compsoc]{IEEEtran}
\ifCLASSOPTIONcompsoc
  \usepackage[nocompress]{cite}
\else
  \usepackage{cite}
\fi
\ifCLASSINFOpdf
\else
\fi
\usepackage{hyperref}
\usepackage{balance}
\usepackage{graphicx}
\usepackage{amssymb}

\usepackage{color}
\usepackage{multirow}
\usepackage{subfigure}
\usepackage{enumerate}
\usepackage{array}
\usepackage{caption}

\usepackage{algorithmicx}
\usepackage{algorithm}
\usepackage{algpseudocode}
\usepackage{xspace}

\newtheorem{lemma}{Lemma}
\newtheorem{corollary}{Corollary}

\newcommand{\todo}[1]{[\textcolor{green}{TODO: #1}]}

\newcommand{\panos}[1]{[\textcolor{magenta}{Panos: #1}]}
\newcommand{\hide}[1]{}
\newcommand{\eat}[1]{}
\newcommand{\stitle}[1]{\vspace{1ex}\noindent\textbf{#1}}
\newcommand{\rev}[1]{#1}
\newcommand{\ext}[1]{\textcolor{black}{#1}}
\newcommand{\revisions}[1]{#1}

\newcommand{\rtree}{\ensuremath{\mathsf{R\mbox{-}tree}}\xspace}
\newcommand{\rstar}{\ensuremath{\mathsf{R\mbox{*-}tree}}\xspace}
\newcommand{\block}{\ensuremath{\mathsf{BLOCK}}\xspace}
\newcommand{\qtree}{\ensuremath{\mathsf{quad\mbox{-}tree}}\xspace}
\newcommand{\mxcif}{\ensuremath{\mathsf{MXCIF\mbox{ }\qtree}}\xspace}

\newcommand{\onelevel}{\ensuremath{\mathsf{1\mbox{-}layer}}\xspace}
\newcommand{\twolevel}{\ensuremath{\mathsf{2\mbox{-}layer}}\xspace}
\newcommand{\twolevelplus}{\ensuremath{\mathsf{2\mbox{-}layer^+}}\xspace}

\newcommand{\refvone}{\ensuremath{\mathsf{Simple}}\xspace}
\newcommand{\refvtwo}{\ensuremath{\mathsf{RefAvoid}}\xspace}
\newcommand{\refvthree}{\ensuremath{\mathsf{RefAvoid^+}}\xspace}

\newcommand{\qatomic}{\ensuremath{\mathsf{queries}\mbox{-}\mathsf{based}}\xspace}
\newcommand{\tatomic}{\ensuremath{\mathsf{tiles}\mbox{-}\mathsf{based}}\xspace}

\begin{document}
%
\title{Two-layer Space-oriented Partitioning for Non-point Data}
%
%
%
%

\author{Dimitrios~Tsitsigkos,~
        Panagiotis~Bouros,~
        Konstantinos~Lampropoulos,~
        Nikos~Mamoulis,~
        \eat{and~}Manolis~Terrovitis
\IEEEcompsocitemizethanks{\IEEEcompsocthanksitem D. Tsitsigkos, K. Lampropoulos and N. Mamoulis are with the Department
of Electrical \& Computer Engineering, University of Ioannina, Greece.\protect\\
E-mail: \{dtsitsigkos, klampropoulos, nikos\}@cse.uoi.gr
\IEEEcompsocthanksitem P. Bouros is with the Institute of Computer Science, Johannes Gutenberg University Mainz, Germany.
E-mail: bouros@uni-mainz.de
\IEEEcompsocthanksitem M. Terrovitis is with the Information Systems Management Institute, Research Center `Athena', Greece.\\
E-mail: mter@athenarc.gr
}
}

\IEEEtitleabstractindextext{%
\begin{abstract}
Non-point spatial objects (e.g., polygons, linestrings, etc.) are
ubiquitous.
We study the problem of indexing non-point objects in memory for range
queries and spatial intersection joins.
We propose a secondary partitioning technique for space-oriented partitioning
indices (e.g., grids), which improves
their performance significantly, by avoiding the generation
and elimination of duplicate results.
Our approach is 
easy to implement and  can
be used by any space-partitioning index
to significantly reduce the cost
of range queries \ext{and intersection joins}.
\rev{In addition, the secondary partitions can be processed
  independently, which makes our method appropriate for distributed
  and parallel indexing.}
Experiments on real datasets
confirm the
advantage of our
approach against alternative duplicate elimination techniques and
data-oriented state-of-the-art spatial indices.
\ext{We also show that our partitioning technique, paired with
  optimized partition-to-partition join algorithms, typically reduces
  the cost of spatial joins by around 50\%.}
\end{abstract}


}

\maketitle

\IEEEdisplaynontitleabstractindextext

%
\IEEEpeerreviewmaketitle

\section{Introduction}
\IEEEPARstart{S}{patial} data management
has been extensively studied, especially for disk-resident data
\cite{2011Mamoulis}.
Nowadays,
in most
applications,
\ext{collections of points or minimum
  bounding rectangles (MBRs) of extended objects}
can easily fit in the memory of
even a commodity machine.
Although a number of scalable systems for spatial data have been developed in the past decade \cite{AjiWVLL0S13,EldawyM15,XieL0LZG16,YuZS19,PandeyKNK18},
the problem of in-memory management of large-scale spatial data has received relatively little attention.

In this paper, we study the problem of indexing \ext{MBRs of}
non-point spatial objects (e.g., polygons, linestrings, etc.) 
in memory, for the efficient
evaluation of
\ext{the {\em filter step} of}
range queries \ext{and spatial intersection joins}. 
Large volumes of non-point data are
ubiquitous, hence, their effective management is always timely.
Besides Geographic Information Systems,
domains that manage big volumes of such data include graphics (e.g., management of huge meshes \cite{Hoppe96}), neuroscience (e.g., building and indexing a spatial model of the brain \cite{PavlovicSHA18}), and location-based analytics (e.g., managing spatial influence regions of mobile users in order to facilitate effective POI recommendations \cite{ChengYKL12}).


\stitle{Motivation.}
Spatial access methods can be divided into two categories; {\em space-oriented partitioning} (SOP) and {\em data-oriented partitioning} (DOP) approaches. Indices of the first category divide the space into {\em spatially disjoint} partitions. 
As a result, objects that overlap with multiple partitions must be replicated (or clipped) in each of them.
DOP methods 
allow the extents of the partitions to overlap and
ensure that their contents are disjoint
(i.e., each object is assigned to exactly one partition).
For disk-resident data, DOP approaches (such as the R-tree
\cite{Guttman84} and its variants) are considered to be the best, because they avoid data replication and they have a balanced structure.
However, SOP approaches (especially grids) are gaining ground due to their efficiency in search and updates in main memory \cite{MokbelXA04,KalashnikovPH04,YuPK05,PapadiasMH05,SidlauskasSCJS09,RayBG14}.
In addition, query evaluation over grids is embarassingly
parallelizable and SOP is widely used in distributed spatial data
management systems \cite{EldawyM15,XieL0LZG16,YuZS19}.

In this paper, we address an inherent problem that SOP indices have: potential duplicate query results.
For example, consider the \eat{six }rectangular objects 
depicted in Figure
\ref{fig:grid}, partitioned using a 4$\times$4 grid.
Some objects 
are assigned to multiple tiles. 
Given a query range
(e.g., $W$), a replicated object (e.g., $r_2$) may be identified as query result multiple times (e.g., at tiles $T_0$, $T_1$, $T_4$, and $T_5$).
The classic, but slow, approach to eliminate duplicates is to hash the query
results and identify duplicates at each bucket \cite{ArefS94}.
The 
state-of-the-art
approach \cite{DittrichS00},
used in most big spatial data management
systems \cite{PandeyKNK18},
computes,
for each query result $r_i$ found in a partition $T$, a
{\em reference point} of the intersection
between $r_i$ and the query window $W$
(e.g., the upper-left corner in Figure~\ref{fig:grid}).
If the reference point is inside $T$, then $r_i$ is reported,
otherwise it is ignored. Since the reference point can only be inside
one partition, no duplicate results are reported.
Although this method avoids hashing, we still have to bear the cost of
finding duplicate results  and
computing the reference point for each of them.

\eat{
\begin{figure}[htb]
\subfigure[grid]{
    \label{fig:grida}
     \includegraphics[width=0.4\columnwidth]{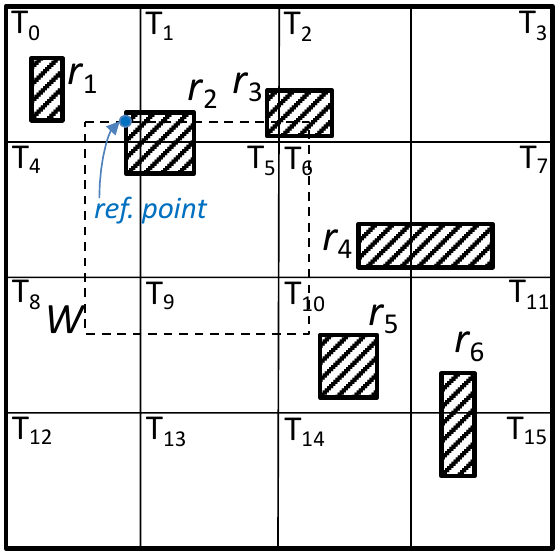}
    }
   \hspace{-0.1in}
  \subfigure[partitioning of rectangles]{
    \label{fig:gridb}
    {\renewcommand{\arraystretch}{1.4}
\begin{tiny}
\begin{tabular}{|@{~}c@{~}|@{~}c@{~}|@{~}c@{~}|}
  \hline
  Tile & primary partitioning & secondary partitioning\\
  \hline
 $T_0$&$\{r_1,r_2\}$&$A=\{r_1,r_2\}$ \\\hline
 $T_1$&$\{r_2,r_3\}$&$A=\{r_3\}$, $C=\{r_2\}$ \\\hline
 $T_2$&$\{r_3\}$&$C=\{r_3\}$\\\hline
 $T_4$&$\{r_2\}$&$B=\{r_2\}$\\\hline
 $T_5$&$\{r_2\}$&$D=\{r_2\}$\\\hline
 $T_6$&$\{r_4\}$&$A=\{r_4\}$\\\hline
 $T_7$&$\{r_4\}$&$C=\{r_4\}$\\\hline
 $T_{10}$&$\{r_5\}$&$A=\{r_5\}$\\\hline
 $T_{11}$&$\{r_6\}$&$A=\{r_6\}$\\\hline
 $T_{15}$&$\{r_6\}$&$B=\{r_6\}$\\\hline
\end{tabular}
\end{tiny}
    }
  }
  \vspace{-0.1in}
  \caption{Example of partitioning and object classes}
  \label{fig: grid}
\end{figure}
}

\begin{figure}[t]
\centering
  \begin{tabular}{c}
    \includegraphics[width=0.55\columnwidth]{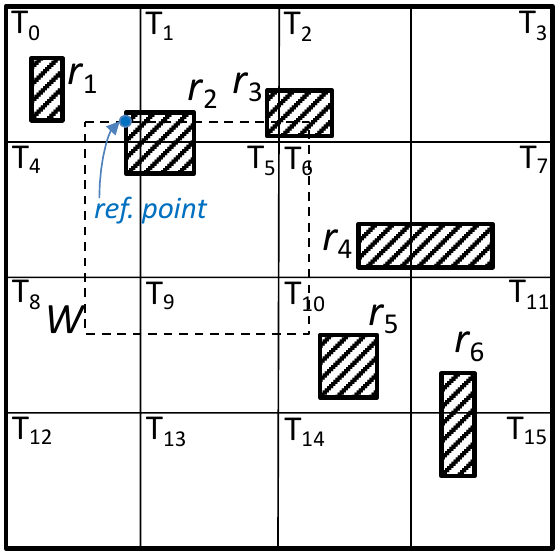}
\\
{\footnotesize
\begin{tabular}{|c|c|c|}
  \hline
  \textbf{tile} &\textbf{primary partitioning} &\textbf{secondary partitioning}\\\hline\hline
 $T_0$&$\{r_1,r_2\}$&$A=\{r_1,r_2\}$ \\\hline
 $T_1$&$\{r_2,r_3\}$&$A=\{r_3\}$, $C=\{r_2\}$ \\\hline
 $T_2$&$\{r_3\}$&$C=\{r_3\}$\\\hline
 $T_4$&$\{r_2\}$&$B=\{r_2\}$\\\hline
 $T_5$&$\{r_2\}$&$D=\{r_2\}$\\\hline
 $T_6$&$\{r_4\}$&$A=\{r_4\}$\\\hline
 $T_7$&$\{r_4\}$&$C=\{r_4\}$\\\hline
 $T_{10}$&$\{r_5\}$&$A=\{r_5\}$\\\hline
 $T_{11}$&$\{r_6\}$&$A=\{r_6\}$\\\hline
 $T_{15}$&$\{r_6\}$&$B=\{r_6\}$\\\hline
\end{tabular}
}
    \end{tabular}
\caption{Example of partitioning and object classes}\label{fig:grid}
\end{figure}

\begin{figure}[t]
\centering
    \includegraphics[width=1.0\columnwidth]{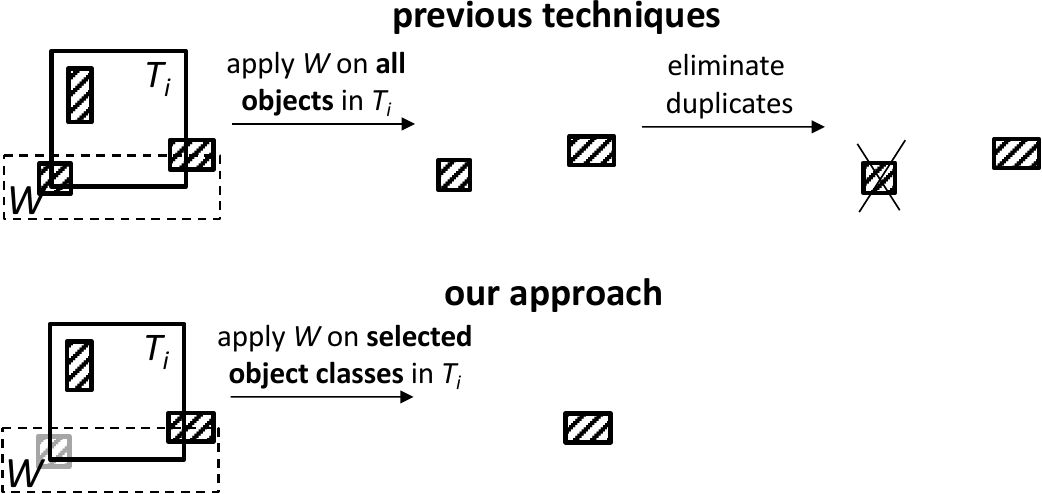}
\caption{Comparing our approach to previous work}\label{fig:illustration}
\end{figure}

\stitle{Contributions.}
\rev{In Section~\ref{sec:index},
we propose a secondary partitioning technique for SOP indices,
which improves their performance
significantly, by avoiding the generation and elimination of duplicate
results.
Our approach
(i) is extremely
easy to implement, (ii) \rev{it can
 be used by any SOP index}, and (iii)
it can be directly implemented in
big spatial 
data management systems
\cite{PandeyKNK18}.
In a nutshell, we divide the objects which are assigned to
each partition $T$ into four classes $A,B,C,D$.
  Objects in class $A$ begin inside $T$ in both dimensions, objects in
  class $B$ start  inside $T$ in dimension $x$ only,
  objects in
  class $C$ start  inside $T$ in dimension $y$ only, and objects in
  class $D$ start before $T$ in both dimensions.
Figure~\ref{fig:grid} (column ``secondary partitioning'')
exemplifies how the objects are divided into
classes.
During query evaluation, for each partition $T$ which intersects the
query range, we access {\em only} the object classes in $T$ that are
guaranteed not to produce
duplicate results.
For example, in tile $T_1$ of Figure~\ref{fig:grid}, we will not
access class $C$, because query $W$ starts before $T_1$ in dimension
$x$; i.e., any object in class $C$ of $T_1$ that intersects $W$ (like $r_2$) should
also intersect $W$ in the previous tile $T_0$.
In Section~\ref{sec:range:classes}, we explain in detail how range queries
are evaluated by our scheme
and show how redundant computations and duplicate checks
can be avoided overall.}
Figure~\ref{fig:illustration} illustrates\ext{, for range queries,} the difference between our
approach and the de-duplication process followed by previous work
\cite{DittrichS00,ArefS94}; while all previous approaches evaluate
queries on {\em all} objects of each partition and then eliminate
possible duplicates, we process only a subset of objects in each
partition that cannot be duplicates
and we do not perform any de-duplication. 

Besides
duplicate avoidance,
our technique can also facilitate reducing 
number of required comparisons per rectangle 
to at most one per dimension (Section~\ref{sec:compreduction}).
\ext{In Sec. \ref{sec:join}, we turn our focus to spatial intersection joins.
  We first discuss join evaluation for two datasets that
are primarily indexed by identical grids.
To avoid duplicate results, we show that, for each tile, it suffices to
evaluate 9 out of the 16 possible joins between the pairs
of secondary partitions in the tile.
We also show how to optimize the join phase of PBSM, by specialized
plane-sweep routines for the different cases of joined sub-partitions
(classes) and by avoidance of redundant comparisons.
Finally, we
investigate  join evaluation when one or both
joined inputs have already been indexed and how to process joins of
datasets that have been partitioned using a different grid.}

In Section~\ref{sec:exps}, we evaluate our proposal experimentally
using large publicly available real datasets and synthetic ones.
Our experiments
show that main-memory grids are superior to
alternative SOP and
DOP indices.
More importantly, we show that when we replace the
state-of-the-art duplicate elimination technique \cite{DittrichS00} by
our secondary partitioning technique, the performance of grid-based 
indexing is improved by up to a few times.

\ext{
  This paper extends of our work in
\cite{TsitsigkosLBMT21} to address spatial intersection joins. For the
interest of space, some content from \cite{TsitsigkosLBMT21} is
omitted from this version (storage decomposition optimization,
handling of distance range queries, refinement step, batch query processing).}
\section{Related Work}
\label{sec:related}

\subsection{Indexing Non-point Spatial Objects}
Spatial queries are typically processed in two steps \cite{2011Mamoulis},
following a \emph{filtering-and-refinement} framework.
During \eat{the }\emph{filtering}\eat{ step}, the query is applied on
the MBRs, which approximate the objects.
During \eat{the }\emph{refinement}\eat{ step}, the exact representations of the candidates are accessed and tested against the query predicate.
Spatial indices are applied in the filtering step; 
hence, they manage MBRs instead of exact geometries.

Depending on the nature of the partitioning, spatial indices are\eat{can be} classified into two classes \cite{OlmaTHA17}.
\emph{Space-oriented partitioning} (SOP) indices divide the space into disjoint partitions and were originally designed for point data.
A grid \cite{BentleyF79}, which divides the space into cells (partitions) using axis-parallel lines,
is the simplest SOP index. 
Hierarchical indices \eat{that fall }in this category are the
kd-tree \cite{Bentley75} and the quad-tree \cite{FinkelB74},\revisions{\cite{MahinHK17}}.
SOP can also be used for non-point objects; in this case,
objects whose extent overlaps with multiple partitions
are
replicated (or clipped) in each of them \cite{Samet90}.

Due to object replication, the same query results may be detected in
multiple partitions \revisions{intersecting the query} and
result deduplication
should be applied \cite{ArefS94}.
Dittrich and Seeger \cite{DittrichS00} avoid duplicate results
by computing a {\em reference point} of the intersection area between
each result and the query range.
If the reference point is inside the partition, then the result is
reported, otherwise it is eliminated.

Indices based on \emph{data-oriented partitioning} (DOP)
allow the extents of the partitions to overlap and
ensure that their contents are disjoint
(i.e., each object is assigned
to exactly one partition); hence, there is no need for result de-duplication.
Variants of the R-tree \cite{Guttman84} (e.g., the R*-tree \cite{BeckmannKSS90})
are the most popular methods in this class.
The R-tree is a disk-based, height-balanced tree, which generalizes the B$^+$-tree in the multi-dimensional space and hierarchically groups object MBRs to blocks.
The CR-tree \cite{KimCK01} is an optimized R-tree for the memory hierarchy. 
BLOCK \cite{OlmaTHA17}
is a main-memory DOP index, which uses a hierarchy of grids.
\revisions{R*-Grove \cite{VuE20} is a spatial partitioning, which builds on the split algorithm of R*-tree to define full blocks and balanced partitions for distributed big data.}

Following a recent trend
on relational data \cite{KraskaBCDP18},
{\em learned indices} for spatial data were proposed 
\cite{WangFX019,Li0ZY020,QiLJK20,NathanDAK20,DingNAK20}.
These indices are not directly comparable to our work, as they 
handle
point data (with no obvious extension to non-point data) and their
\eat{primary }goal is to minimize the I/O cost.

\ext{
\subsection{Spatial Joins}
A plane-sweep algorithm \cite{PreparataS85,ArgePRSV98} which computes rectangle intersections, 
is the most common approach for in-memory \eat{processing small }spatial joins. Brinkhoff et al. \cite{BrinkhoffKS93} present an {\em forward sweep} variant, which does not use any data structures.
For datasets too large to fit in memory, data partitioning has been used in a divide-and-conquer fashion, such that inputs are split into smaller subsets which can  be then joined fast in memory. A partition from input $R$ is then joined with a partition from $S$ only if their MBRs intersect. Partitioning-based approaches are divided in two categories.
\emph{Single-assignment, multi-join} (SAMJ) methods assign each object to exactly one partition, similar to DOP; then, a partition from one input may
be joined with multiple partitions from the other.
The R-tree Join algorithm \cite{BrinkhoffKS93} is a classic SAMJ approach, where both inputs are indexed by an R-tree. 

On the other hand, \emph{multi-assignment, single-join} (MASJ) methods, inspired by SOP, use identical and disjoint spatial partitions for both datasets. An \eat{input }object is assigned to every partition it spatially intersects and each partition from one input is joined with exactly one partition from the other.
\emph{Partition-based Spatial Merge Join} (PBSM) \cite{PatelD96}, the state-of-the-art MASJ method \cite{NobariQJ17,TsitsigkosBMT19}, divides the space by a regular grid\eat{ and assigns objects from both input collections to all tiles that spatially overlap them}.
For each spatial partition, PBSM accesses the objects from both inputs and performs the
small join in memory (e.g., using plane-sweep).
Duplicate join results are eliminated, using the reference point technique \cite{DittrichS00}). Other MASJ approaches include \emph{Spatial Hash Join} \cite{LoR96} and \emph{Scalable Sweeping-Based Spatial Join} \cite{ArgePRSV98}.

In-memory methods
also utilize partitioning to accelerate the join computation.
TOUCH \cite{NobariTHKBA13} first bulk-loads an R-tree for one of the inputs, using STR packing \cite{LeuteneggerEL97}. Then, all objects from the second input are assigned to buckets corresponding to the non-leaf nodes of the tree.
Finally, each bucket is joined with the subtree rooted at the corresponding node, as in \cite{MamoulisP03}.
A comparison of spatial join algorithms for in-memory data \cite{NobariQJ17} shows that PBSM and TOUCH perform best.
Tauheed et al. \cite{TauheedHA15} suggest an analytical model for configuring the grid of PBSM-like join processing.

}

\subsection{Parallel and Distributed Data Management}
With the advent of Hadoop, research on spatial data management has
shifted to developing distributed systems for spatial data \cite{CarySHR09,AjiWVLL0S13,EldawyM15,YouZG15,XieL0LZG16,YuZS19}.
Spatial data in {\em Hadoop-GIS} \cite{AjiWVLL0S13} are partitioned
using a hierarchical grid, wherein high density tiles are split to
smaller ones.
The nodes of the cluster share a {\em global tile index} which can be used to find
the HDFS files where the contents of the tiles are stored.
Spatial queries are
implemented as MapReduce workloads. 
{\em SpatialHadoop} \cite{EldawyM15},
offers different options for partitioning
(i.e., grid based, R-tree based, quad-tree based, etc.)
The Master node holds a global spatial index for the MBRs of each of
the HDFS file blocks.
A local index is built at each physical partition and used by map tasks.

Spark-based implementations of spatial data management systems
\cite{YouZG15,XieL0LZG16,YuZS19},\revisions{\cite{EldawyHGSSSSSVZ21}}
apply similar
partitioning approaches. The main difference to Hadoop-based systems
is that data, indices, and intermediate results are shared in
the memories of all nodes in the cluster 
as RDDs.
Unlike SpatialSpark \cite{YouZG15}, GeoSpark (now, Apache Sedona)
\cite{YuZS19} and \revisions{Beast \cite{EldawyHGSSSSSVZ21}} which are
built on top of Spark, Simba \cite{XieL0LZG16} has its own native
query engine and query optimizer, but 
does not support non-point geometries.
Pandey et al. \cite{PandeyKNK18}
conduct a comparison between big spatial data analytics systems.

\ext{Distributed spatial data management systems focus on data
partitioning and not on query evaluation
at each partition. Emphasis is given on scaling
out,
rather than reducing the
computational cost per node.
On the other hand, we focus on in-memory spatial data management and
scaling up,
by reducing the computational cost of spatial query evaluation\eat{ and exploiting
multi-core parallelism}.
Still, our ideas are readily applicable in share-nothing
systems, where each spatial partition is independent
and does not need input from others to produce results.
}

\section{Two-layer Spatial Partitioning}
\label{sec:index}
In this section, we present our secondary partitioning approach for
SOP spatial indices.
Although our approach can be used in any SOP index,
we will present
it in the context of a \rev{regular} grid,
which
divides the space into $N\times M$ disjoint spatial partitions, called
{\em tiles}.
%
An object $o$ is assigned to a tile $T$
iff $MBR(o)$ and $T$ intersect.
For each tile $T$, we keep a list of (MBR, object-id) pairs
that are assigned to $T$.
\rev{The actual geometries of objects are stored separately and
  retrieved  on-demand, based on their id's.}
If the spatial distribution of objects is not uniform
and there are many empty tiles,
to save memory,
we use a hash-table to access non-empty tiles based on their
coordinates.



\stitle{Secondary Partitioning.}
We propose that the (MBR, object-id) pairs
at each tile are further divided into
four classes $A$, $B$, $C$, and $D$
(which are physically stored separately in memory).
Let $r.x=[r.x_l, r.x_u]$ be the projection of MBR $r$ on the
$x$ axis and $r.y=[r.y_l, r.y_u]$ $r$'s $y$-projection.
Now, consider an MBR $r$ which is assigned to tile $T$.
\begin{itemize}
\item
  $r$ belongs to class $A$, if for every dimension
    $d\in \{x,y\}$, the begin value $r.d_l$ of $r$ falls into projection
    $T.d$, i.e., if $T.d_l\le r.d_l$.
  \item
    $r$ belongs to class $B$ if $r.x$
       begins inside $T.x$ and $r.y$ begins before
       $T.y$, i.e.,  if $T.x_l\le r.x_u$ and $T.y_l> r.y_l$.
     \item
       $r$ belongs to class $C$ if $r.x$
       begins before $T.x$ and $r.y$ begins inside
       $T.y$, i.e.,  if $T.x_l> r.x_l$ and $T.y_l\le r.y_l$.
     \item
       $r$ belongs to class $D$ if both its $x$- and $y$-projections
       begin before $T$, i.e., if $T.x_l> r.x_l$ and $T.y_l> r.y_l$.
\end{itemize}

Figure \ref{fig:classes} illustrates examples of rectangles in a tile
$T$ that belong to the four different classes.%
\footnote{We conventionally assume that the $x$ dimension is from left to right and the $y$ dimension is from top to bottom.}
During data partitioning,
when a rectangle $r$ is assigned to a tile $T$,
we identify its class and place it to the corresponding division.
Note that a rectangle can belong to class $A$
of just one tile, while it can belong to other classes (in other
tiles) an arbitrary number of times.
We denote the secondary partitions of tile $T$ 
which store the MBRs of classes $A$, $B$,
$C$, and $D$, by $T^A$, $T^B$,
$T^C$, and $T^D$, respectively.
\rev{Table \ref{table:notations} summarizes the notation used frequently in
the paper.}

\begin{figure}[t]
  \centering
  \includegraphics[width=0.99\columnwidth]{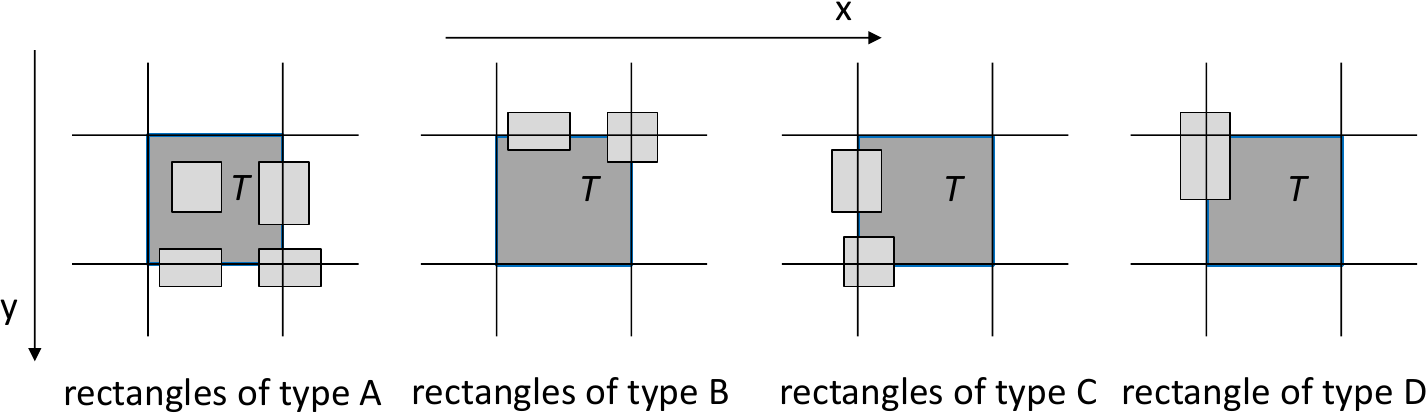}
  \caption{The four classes of rectangles assigned to a tile $T$.}
  \label{fig:classes}
\end{figure}
\eat{
\begin{figure*}[t]
  \centering
  \includegraphics[width=0.8\textwidth]{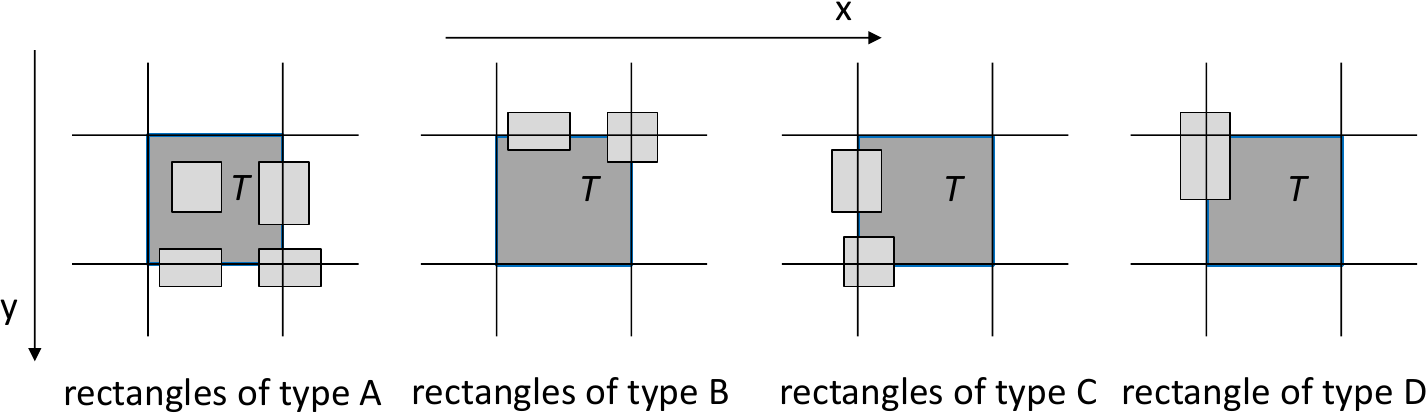}
  \caption{The four classes of rectangles assigned to a tile $T$.}
  \label{fig:classes}
\end{figure*}
}

\begin{table}[t]
\caption{\rev{Table of notations}\label{table:notations}}
\centering
\footnotesize
\begin{tabular}{|c|p{6cm}|}
\hline
\rev{Notation} &\rev{Description}\\ %
\hline \hline
  \rev{$W$} &\rev{query window} \\\hline
   \rev{$r.d=[r.d_l, r.d_u]$} &\rev{projection of rectangle $r$ at dimension $d\in  \{x,y\}$}\\\hline
  \multirow{2}{*}{\rev{$T^X$}} &\rev{secondary partition of tile $T$
                                 holding MBRs in class $X\in\{
        A,B,C,D\}$} \\\hline
  \rev{$prev(T,d)$} &\rev{previous tile to $T$ in dimension $d\in \{x,y\}$}\\\hline
\end{tabular}
\end{table}

\eat{
\subsection{Evaluating queries over a simple grid}
\label{sec:index:grid}
We now discuss in more detail how this simple indexing scheme can be
used to evaluate rectangular range queries and expose its limitations.
We first introduce some notation that will also be useful
when we discuss our solution.

Recall that each MBR $r$ can be represented by an interval of
values at each dimension.
Let $r[i]=[r[i][0], r[i][1]]$ be the projection of rectangle $r$ on the $i$-th axis.
For example, in the 2D space, $r[0][1]$ denotes the
upper bound of rectangle $r$ on dimension $0$ (i.e., the $x$-axis).
Similarly, we use $T[i]=[T[i][0], T[i][1]]$ to denote the projection of a tile $T$ to the
$i$-th dimension.
Given a tile $T$ and a dimension $i$, we use $prev(T,i)$ to denote the
tile $T'$ which is right before $T$ in dimension $i$ and has exactly
the same projection as $T$ in the other dimension(s).
For example, in Figure \ref{fig:index1}, $T_b=prev(T_a,0)$.
$prev(T,i)$ is not defined
for tiles $T$ which are in the first column (for $i=0$)
or row (for $i=1$) of the grid.

Given a range
query window $W$,
a tile that does not intersect $W$ does not contribute any
results to the query. Specifically, the only tiles $T$ that may contain
query results are those for which $T[i][1]\ge W[i][0]$ and $T[i][0]\le
W[i][1]$ at every dimension $i$ and can easily be enumerated after
finding the tiles $T_s$ and $T_e$, which contain
$W[0][0]$ and $W[1][1]$, respectively.%
\footnote{$T_s$ and $T_e$ can be found in $O(1)$ by algebraic
  calculations if the grid is uniform.} 
Figure \ref{fig:index1} illustrates a window query $W$ in lightgrey color
and its four corner points $W[0][0]$, $W[0][1]$, $W[1][0]$, $W[1][1]$.
The tiles which are relevant to $W$ are between (in both dimensions)
the two tiles $T_s$ and $T_e$.%
\footnote{We conventionally assume that the $x=0$ dimension is from
      left to right and the $y=1$ dimension is from top to bottom.}

For each tile which is totally covered by the query range
in at least one dimension (e.g., $T_a$ in dimension $0$),
we know that the objects in it certainly intersect $W$ in that dimension. For a
tile $T$ that partially overlaps with $W$ in both dimensions (e.g., $T_b$), we need to
iterate through its objects list to verify their intersection with $W$.
We first check whether the MBR of the object intersects $W$ and then
we might have to verify with the exact geometry of the object at a
refinement step.

An important issue is that neighboring tiles may intersect $W$ and
also contain the same object $o$. In this case, $o$ will be reported
more than once, so we need an approach for handling these duplicates.
For example, in
Figure \ref{fig:index1}, object $o_1$ could be
reported both by $T_a$ and by $T_b$.
A solution to this problem
is to report an object $o$ only at the tile which is before all tiles
(in both
dimensions) where $o$ is found to intersect $W$.
For example, in
Figure \ref{fig:index1}, $o_1$ is 
reported by $T_b$ only, which is before $T_a$.
An easy approach to perform this test is to compute the 
intersection between the query window and the rectangle and report the
result only if a {\em reference point} of the intersection (e.g., the
smallest values of the intersection in all dimensions)
is included in the tile \cite{DittrichS00}.
While this solution prevents reporting duplicates, 
it requires extra computations and comparisons
and it is unclear how to apply it 
for
non-rectangular range queries.
An alternative and more general (but more expensive)
approach is to add the results from all tiles in a hash table, which
would prevent the same rectangle from being added multiple times.


}
\section{Range Query Evaluation}
\label{sec:range}
\rev{
In this section, we show how the secondary partitions  
at each tile $T$
can be used to
avoid the generation and
elimination of duplicate query results.}
%
\eat{We first consider rectangular range queries
$W$ (window queries).
For now, we focus on the {\em filtering step} of the query, i.e., the
objective is to just find the object MBRs which intersect $W$.
The refinement step will be discussed in Section \ref{sec:refine}.
}
\ext{We discuss the case of rectangular queries $W$ (window queries) and focus on the filtering step of the evaluation; details on non-rectangular queries and on the refinement step can be found in \cite{TsitsigkosLBMT21}.}

First, the tiles 
which intersect $W$
in a $N\times M$ \rev{regular} grid
\rev{can be found in $O(1)$ time}, by algebraic
operations. Specifically, assuming that tile $T_{i,j}$ is at the
$i$-th row and at the $j$-th column of the grid, the tiles which
intersect $W$ are all tiles $T_{i,j}$,
for which $\lfloor W.x_l/N
\rfloor\le i \le \lfloor W.x_u/N \rfloor$ and
$\lfloor W.y_l/M \rfloor \le j \le \lfloor W.y_u/M \rfloor$.
%
We now explain in detail, for each tile $T$ that intersects $W$, which
classes of rectangles should be accessed and which computations are
necessary for determining whether each rectangle $r$ intersects $W$.

\subsection{Selecting relevant classes}
\label{sec:range:classes}
\begin{figure}[t]
  \centering
\includegraphics[width=\columnwidth]{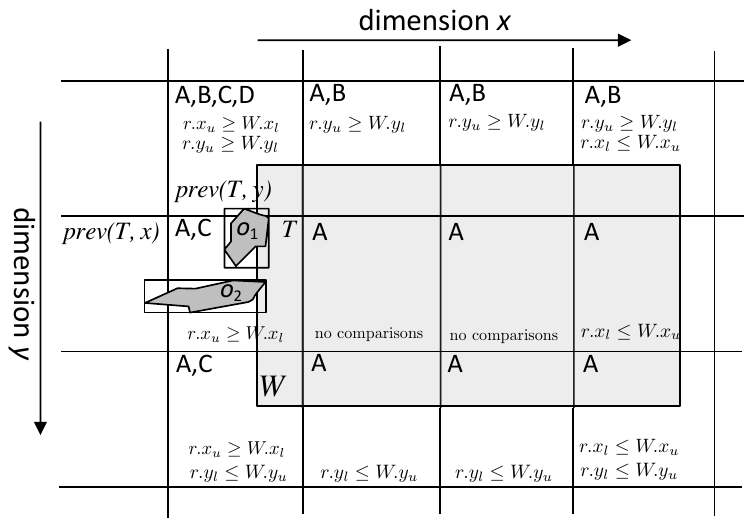}
\caption{Examples of object classes and comparisons}
\label{fig:ABCDexam}
\end{figure}

For a tile $T$, let $prev(T,d)$ denote the
tile which is right before $T$ in dimension $d$ and has exactly
the same projection as $T$ in the other dimension.
For example, in Figure~\ref{fig:ABCDexam}, $prev(T,x)$ (resp. $prev(T,y)$) is the tile
right before $T$ in dimension $x$ (resp. $y$).
Given a window query $W$, 
the following lemmas determine the
classes of rectangles in $T$ which
should be disregarded,
because they can only produce duplicate results.


\begin{lemma}
  \label{lemma:filteringx}
  If the query range $W$ intersects tile $T$ and $W$ starts before $T$ in dimension $x$,
  then secondary partitions  $T^C$ and $T^D$ should be disregarded.
\end{lemma}

{\em Proof. }
  Consider a rectangle $r$ in class $C$ or class $D$ of tile $T$, i.e., $r \in
  T^C$ or $r \in T^D$. Rectangle $r$ should also be assigned to the previous tile
  $prev(T,x)$ to $T$ in dimension $x$, because it belongs to class $C$
  or $D$ of $T$.
  If $r$ intersects $W$ in $T$, then $r$ should also intersect $W$ in
  $prev(T,x)$, because $W$ also starts before $T$ in dimension $x$.
  Hence, examining and reporting $r$ in tile $T$ would produce a duplicate, since
  the same result can also be identified in tile
  $prev(T,x)$. \hfill $\square$
  
\begin{lemma}
  \label{lemma:filteringy}
  If $W$ intersects tile $T$ and $W$ starts before $T$ in dimension $y$,
  then secondary partitions  $T^B$ and $T^D$ should be disregarded.
\end{lemma}

Lemma~\ref{lemma:filteringy} can be proved by replacing $x$ by $y$ and
$C$ by $B$ in the proof of Lemma~\ref{lemma:filteringx}.
The two lemmas are combined to exclude all classes $B$, $C$, and $D$ if  $W$ starts before $T$ in both dimensions. 
To illustrate the lemmas, consider
tile $T$ in Figure~\ref{fig:ABCDexam}.
In addition, consider the MBRs of objects $o_1$ and $o_2$, which belong to
secondary partitions $T^B$ and $T^C$, respectively.
$MBR(o_1)$ should be ignored when processing $T$ because it belongs to
class $B$ and $W$ starts before $T$ in dimension $y$ (Lemma~\ref{lemma:filteringy}). Indeed, $MBR(o_1)$ intersects $W$ also
in tile $prev(T,y)$ which is right above $W$.
On the other hand, $W$ does not start before $T$ in dimension
$x$, i.e., Lemma~\ref{lemma:filteringx} does not apply for tile $T$.
This means that $MBR(o_2) \in T^C$ will be found to 
intersect $W$.
Figure~\ref{fig:ABCDexam} shows, in the top-left corner of each tile
$T$ intersected by $W$,
the object classes in $T$ that should be examined (the
remaining classes can be disregarded).
Observe that we have to consider all objects in just one tile (the one
containing point $(W.x_l,W.y_l)$). For the majority of tiles, we only
have to examine secondary partition $T^A$.

\subsection{Minimizing \rev{the number of} comparisons}\label{sec:compreduction}
We now turn our attention to {\em minimizing the  number of comparisons} needed for each
secondary partition that {\em  has to} be checked
(i.e., those not eliminated by Lemmas
\ref{lemma:filteringx} and \ref{lemma:filteringy}).
\rev{
  For a rectangle $r$ in a tile $T$ to intersect the query
window $W$, $r.x$ should intersect $W.x$ and
$r.y$ should intersect $W.y$.
Hence, to test whether $x$ intersects $W$,
we need at most four comparisons (i.e.,
$r$
and $W$ do not intersect, iff $r.x_u<
W.x_l$ or $r.x_l> W.x_u$ or $r.y_u< W.y_l$ or $r.y_l> W.y_u$).}

A direct observation that saves comparisons is that,
if a tile $T$ is covered by the window $W$ in a dimension $d$,
then we do not have to perform intersection tests in dimension $d$
for all rectangles in the relevant secondary partitions in $T$.
In the example of Figure \ref{fig:ABCDexam}, we need to examine
partitions $T^A$ and $T^C$ of tile $T$ (Lemma \ref{lemma:filteringy}).
For each rectangle $r$ in these partitions, we only have to verify if
projections $r.x$ and $W.x$ intersect, because $r.y$ and $W.y$
definitely intersect (since $T.y$ is covered by $W.y$).

For the dimension(s) where $T$ is not covered by $W$,
the following lemmas can be used to further reduce the necessary comparisons.


\begin{lemma}
  \label{lemma:comp1}
  If $W$ ends in tile $T$ and starts before $T$ in dimension $d$, then
  for a rectangle $r\in T$, $r$ intersects $W$ in dimension $d$ iff
  $r.d_l \le W.d_u$. 
\end{lemma}

Symmetrically, we can show:
\begin{lemma}
  \label{lemma:comp2}
  If $W$ starts in tile $T$ and ends after $T$ in dimension $d$, then
  for a rectangle $r\in T$, $r$ intersects $W$ in dimension $d$ iff $r.d_u\ge W.d_l$. 
\end{lemma}

For example, in tile $T$ of Figure \ref{fig:ABCDexam},
we only have to test intersection in
dimension $x$, as already explained.
The intersection test can be
reduced to a simple comparison,
i.e., $r$ intersects $W$ iff $r.x_u\ge W.x_l$, due to Lemma \ref{lemma:comp2}.
To demonstrate the impact of Lemmas \ref{lemma:comp1} and
\ref{lemma:comp2},
in each tile of the figure, we show the necessary comparisons.
For the two tiles in the center,
no comparisons are required because all MBRs (in class A) are
guaranteed to intersect $W$. For the remaining two tiles, which
intersect the border of $W$,
we only have to perform at most one comparison per dimension,
because $W$ either starts or ends at these tiles (and some of these
tiles are totally covered by $W$ in one dimension).
Contrast this to the four comparisons required in the general case for testing
whether two rectangles (e.g., $r$ and $W$) intersect.
Therefore, for range queries that cover multiple tiles, we have:

\begin{corollary}
  \label{cor:comp}
  For a window query $W$ that intersects more than one tile per
  dimension,
  at most two comparisons per rectangle in each relevant tile are required.
\end{corollary}

\eat{
\subsection{Storage decomposition}\label{sec:decomposition}

\rev{Conventionally, each MBR $r$ is stored as a
  quintuple   $\langle id, r.x_l, r.x_u, r.y_l, r.y_u\rangle$.
To further reduce the query cost and 
improve the
data access locality, we propose
the representation of each MBR $r$ by
 four pairs:
$\langle r.x_l, id\rangle$, $\langle r.x_u,
id\rangle$, $\langle r.y_l, id\rangle$, $\langle r.y_u, id\rangle$,
following
the Decomposition Storage Model (DSM) \cite{CopelandK85,StonebrakerABCCFLLMOORTZ05}.
Hence, for each of the
secondary
partitions $T^X\in \{T^A$, $T^B$, $T^C$, $T^D\}$, we can define four
decomposed tables $L^X_{x_l}, L^X_{x_u},
L^X_{y_l}, L^X_{y_u}$, which store the four pairs of
each rectangle in $T^X$, respectively.}
The tables are sorted by their first column and used to evaluate fast
queries on tiles, where just one endpoint of each MBR needs to be compared
(according to Lemmas \ref{lemma:comp1} and
\ref{lemma:comp2}).

In particular, for each tile $T$ satisfying Lemma \ref{lemma:comp1} in
dimension $d$, we can perform {\em binary search} on \rev{each of its relevant 
tables $L^X_{d_l}$, having the $\langle r.d_l, id\rangle$ tuples,}
to find the largest  $r.d_l$, which
satisfies $r.d_l \le W.d_u$. All rectangles in the table up to this
value are guaranteed to satisfy the condition and can be reported
without any comparison.%
\footnote{Alternatively, we can scan from the beginning of the table
  until the condition is violated.}
Symmetrically, we can reduce the comparisons for rectangles in a tile
$T$, which  satisfies Lemma \ref{lemma:comp2}, by taking advantage of \rev{the
sorted tables $L^X_{d_u}$}.
For example, for the tile $T$ in Figure \ref{fig:ABCDexam}, we
only have to access and perform binary search to tables $L^A_{x_u}$
and $L^C_{x_u}$, which store the $\langle r.x_u, id\rangle$
decompositions of the rectangles in secondary partitions
$T^A$ and $T^C$, respectively.
If we have to perform two comparisons in a tile (e.g., 
$r.x_u\ge W.x_l$ and $r.y_u\ge W.y_l$), we choose
one of the two relevant decomposed tables (e.g., $L_{x_u}$ or $L_{y_u}$)
to perform the search; then, for each qualifying rectangle according to
the selected comparison, we verify the other comparison by accessing
the entire MBR.
We select the table in the dimension which is covered the least by
$W$, in order to minimize the necessary verifications.

Finally, we observe that, for some object classes, it is not necessary
to store all decompositions. For example, the only possible
comparisons that can be applied to rectangles of class $D$ are  
$r.x_u\ge W.x_l$ and $r.y_u\ge W.y_l$, because all MBRs of class $D$
start before the tile in both dimensions and they are only compared
with $W$ in the tile that includes the start point of $W$ in both
dimensions (Lemma \ref{lemma:comp2}). Hence, we only need to keep
tables $L^D_{x_u}$ and $L^D_{y_u}$ for each secondary partition $T^D$.
Overall,
we can reduce the storage requirements for the decomposed tables as
shown in Table \ref{tab:decompclasses}.

\begin{table}
\centering
\renewcommand{\arraystretch}{1.3}
\caption{Required decomposed tables for each secondary partition}
\label{tab:decompclasses}
\begin{tabular}{|l|c|}\hline
\textbf{partition}				&\textbf{required tables}\\\hline\hline
$T^A$ & $L^A_{x_l}$, $L^A_{x_u}$, $L^A_{y_l}$, $L^A_{y_u}$\\\hline
$T^B$ & $L^B_{x_l}$, $L^B_{x_u}$, $L^B_{y_u}$\\\hline
$T^C$ & $L^C_{x_u}$, $L^C_{y_l}$, $L^C_{y_u}$\\\hline
$T^D$ & $L^D_{x_u}$, $L^D_{y_u}$\\\hline
\end{tabular}
\end{table}

The decomposed data representation not only reduces the number of
comparisons but also accesses only the necessary data for each
verified comparison. In particular, rentangle coordinates which are
not relevant to the required verification are not accessed at all,
while in a record-based representation irrelevant data are fetched to
the memory cache. On the other hand, the decomposed representation
requires additional storage and is more expensive to update (unless
a batch update strategy is employed); hence, it
is mostly appropriate for indexing static spatial object collections. 
}

\subsection{Overall approach}

Algorithm \ref{algo:windowquery} describes the steps of window query evaluation.
Given a window $W$,
we first identify the tiles $\mathcal{T}$ that intersect $W$ by simple algebraic
operations, as discussed in the beginning of this section.
\rev{Then, for each tile $T\in \mathcal{T}$, we
  identify the secondary partitions $\mathcal{P}_T$ that would not
  produce duplicates, using
  Lemmas \ref{lemma:filteringx} and \ref{lemma:filteringy}.
  For each such secondary partition $T^X$, we find all rectangles that
  intersect $W$, by applying the techniques of Section~\ref{sec:compreduction} to reduce the necessary computations.
}
Note that query evaluation at each tile $T$ (and each secondary
partition)
is {\em totally independent} from others and, hence,
it is embarrassingly parallelizable.

\begin{algorithm}[t]
\begin{algorithmic}[1]
\small
\Require grid $\mathcal{G}$, query window $W$
\State $\mathcal{T}$ = tiles in $\mathcal{G}$ that intersect $W$
\For{each tile $T\in \mathcal{T}$}
\rev{   \State $\mathcal{P}_T$ = sub-partitions of $T$ relevant to
   $W$ \Comment Lemmas \ref{lemma:filteringx} \& \ref{lemma:filteringy}
   \For{each sub-partition $T^X \in \mathcal{P}_T$}
      \State find all $r\in  T^X$ that intersect $W$ \Comment
      Section~\ref{sec:compreduction}
   \EndFor
\EndFor}
\end{algorithmic}
\caption{Window query evaluation (filtering step)}
\label{algo:windowquery}
\end{algorithm}

\rev{
Although we focus on indexing 2D MBRs in this paper, our
secondary partitioning scheme can directly be used for minimum bounding boxes
(MBBs) of arbitrary dimensionality $m$. In a nutshell, we need $2^m$
classes to re-partition an $m$-dimensional tile $T$, which indexes
$m$-dimensional MBBs. For each tile $T$, if MBB $r$ intersects $T$,
there are two cases for each dimension
$d$: either $r$ begins inside $T$ ($r.d_l\ge T.d_l$)
or before $T$ ($r.d_l< T.d_l$). Hence, there are $2^m$ cases (classes)
in total.
Lemmas \ref{lemma:filteringx} and \ref{lemma:filteringy} can be
generalized to a lemma that prunes all classes corresponding to cases
of MBBs that begin before $T$ in each dimension $d$, if $W$ begins
before $T$ in that dimension.
Lemmas \ref{lemma:comp1} and \ref{lemma:comp2} apply to any
dimensionality.}

\eat{
\subsection{Non-rectangular ranges}\label{sec:disk}

Window queries are the most popular range
queries. Still, not all query ranges are rectangular.
A characteristic non-rectangular range query
is the {\em disk} (or {\em distance}) range 
query, where the objective is to find all objects with (minimum)
distance to a given query point $q$ at most $\epsilon$.
To evaluate a disk query on our two-layer partitioned dataset, we
apply a similar method to Algorithm \ref{algo:windowquery};
we first find the set of tiles $\mathcal{T}$ that intersect with the disk (using
algebraic/trigonometric operations) and then
find the objects in them that satisfy the query predicate.
As in window queries,
for each tile $T\in \mathcal{T}$,
we check whether $prev(T,d)$ in each dimension $d$ is
also in $\mathcal{T}$. If yes, then we
disregard the corresponding class of rectangles in $T$. Hence, if
$prev(T,x)\in \mathcal{T}$, then classes $B$ and $D$ are disregarded, whereas if
$prev(T,y)\in S$, then classes $C$ and $D$ are disregarded.
Figure \ref{fig:disk} shows an example of a disk query centered at
q. The tiles which intersect the disk are shown by different patterns
depending on the classes of rectangles in them that have to be
checked. For example, in tile $T_5$ all four classes will be examined
(we call $T_5$ an $ABCD$ tile, in the context of the disk query).
Note that for the majority of tiles which intersect the disk range, we
only have to examine rectangles in class $A$.

\begin{figure}[t]
\centering
  \includegraphics[width=\columnwidth]{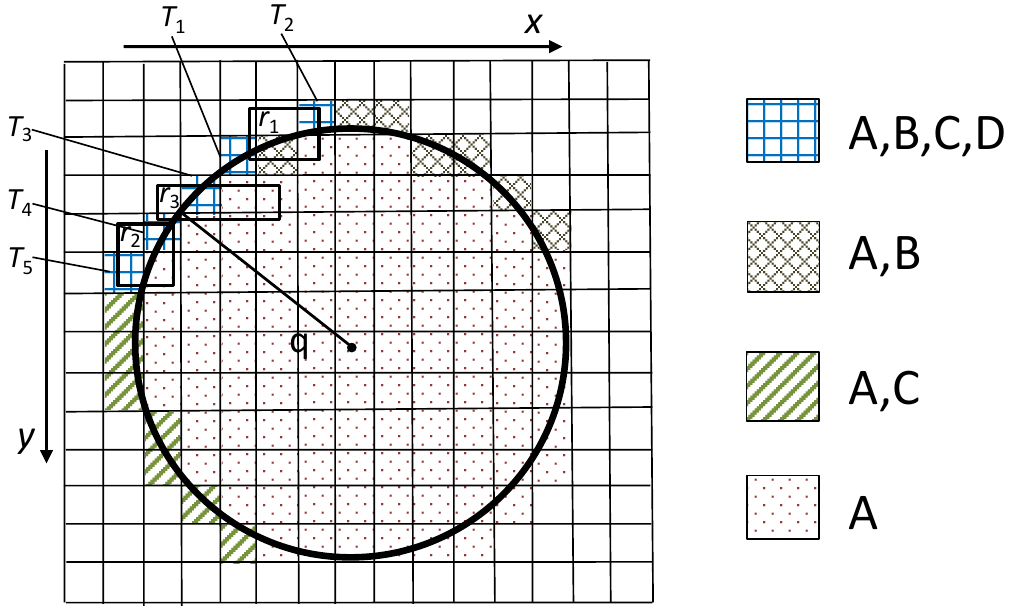}
  \caption{Example of disk query evaluation}
  \label{fig:disk}
\end{figure}

A subtle point here is that if we simply examine all
rectangles in the classes that correspond to each tile, we may end up
examining duplicates. For example, consider rectangle $r_1$, which
will be examined in both tiles $T_1$ (in class $B$) and $T_2$ (in class
$C$). To avoid such duplicates, for each rectangle in an $ABCD$ tile $T$, 
if the tile is closer to $q$ in the $y$-dimension compared to the
$x$-dimension, we ignore rectangles $r$ in classes $C$ and $D$,
for which $r.y_u>T.y_u$ (these will be handled in another tile).
The case where $T$ is closer to $q$ in the $x$-dimension is handled symmetrically. 

For tiles which are totally covered by
the disk range, we do not verify any distances between the objects 
assigned to them and $q$, as these are guaranteed to be query results.
Distance verification only has to be performed for objects in tiles
which partially intersect the disk.

The method described above for disk queries can be
generalized for any non-rectangular query.
We first find the set of tiles $\mathcal{T}$ which
intersect the query range. Then, for each tile $T\in S$, we determine which
classes of objects need to be examined (i.e., exclude classes that
would produce duplicates). For each tile which is totally
covered by the query region, we just report its contents in the
relevant classes as results and for the remaining tiles we conduct an
intersection test for each rectangle before determining whether it is
a result.
}

\ext{
\section{Spatial Join Evaluation}
\label{sec:join}
We now turn our focus to the evaluation of spatial
intersection
joins. First, we discuss how our two-layer partitioning can be adopted
to natively compute
a spatial join,
while avoiding the generation and elimination of duplicate results. 
Then, we elaborate on possible join strategies 
which use two-layer partitioning in different fashions.
\revisions{For illustration purposes, we discuss the above for regular grids; in Section~\ref{sec:join:sops}, we consider \eat{the case of }other SOPs, e.g., the quad-tree.}

\subsection{Two-layer Partitioning Join}
\label{sec:join:mini-joins}
Assume that both \eat{join }input\eat{ dataset}s $R$, $S$ are indexed by our two-layer partitioning\eat{ scheme, specifically \twolevel} 
with identical grids.
In the next subsection, we discuss\eat{elaborate on} the origins and \eat{the }details of such a setting, i.e., whether both or one of $R$, $S$ are (re)-indexed online.
Under this setting, we can build upon the join phase of the PBSM algorithm
\cite{PatelD96} and apply a partition-to-partition join for each
pair of partitions from $R$, $S$ from\eat{that correspond to} the same tile.
\revisions{Assuming that two partitions of size $n$, $m$ are joined, the join cost using 
plane-sweep is $O((n+m)\log(n+m))$, based on \cite{BrinkhoffKS93,ArgePRSV98}.}

\subsubsection{The Mini-joins Breakdown}
\label{sec:join:mini-joins:breakdown}
Given a tile $T$, let $R_T$ and $S_T$ be the partitions containing the object rectangles from
datasets $R$ and $S$, respectively,  that are assigned to $T$.
$R_T$ is divided into rectangle classes $R^A_T$, $R^B_T$, $R^C_T$,
and $R^D_T$, according to our two-layer partitioning scheme discussed in Section~\ref{sec:index}.
Similarly, $S_T$ is divided into $S^A_T$, $S^B_T$, $S^C_T$,
and $S^D_T$. Hence, the spatial join $R_T\bowtie S_T$ can now be
decomposed into $4 \cdot 4 = 16$ joins between classes of rectangles,
i.e.,
$R^A_T \bowtie S^A_T, R^B_T \bowtie S^B_T, R^B_T \bowtie S^C_T, \dots, R^B_T \bowtie
S^A_T,\dots$. We call these class-to-class joins, \emph{mini-joins}. Figure~\ref{fig:minijoins} exemplifies the decomposition of the partition-to-partition join inside a tile $T$ into the 16 mini-joins.

\begin{figure}[t]
\centering
  \includegraphics[width=\columnwidth]{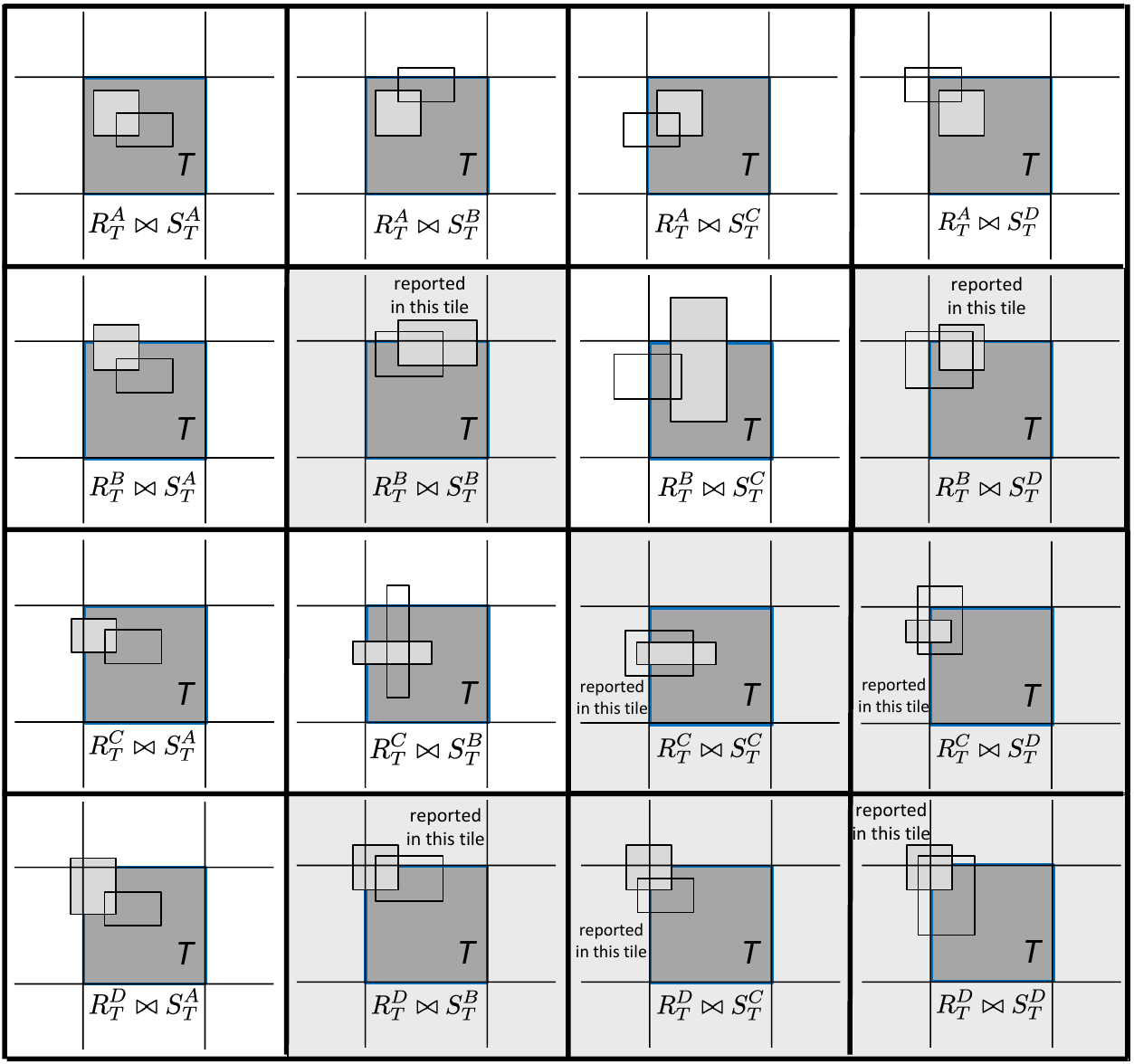}
  \caption{Decomposition of tile $T$'s join into 16 mini-joins; $R$ rectangles filled in light gray color.}
  \label{fig:minijoins}
\end{figure}
\eat{
\begin{figure*}[t]
\centering
  \includegraphics[width=0.8\textwidth]{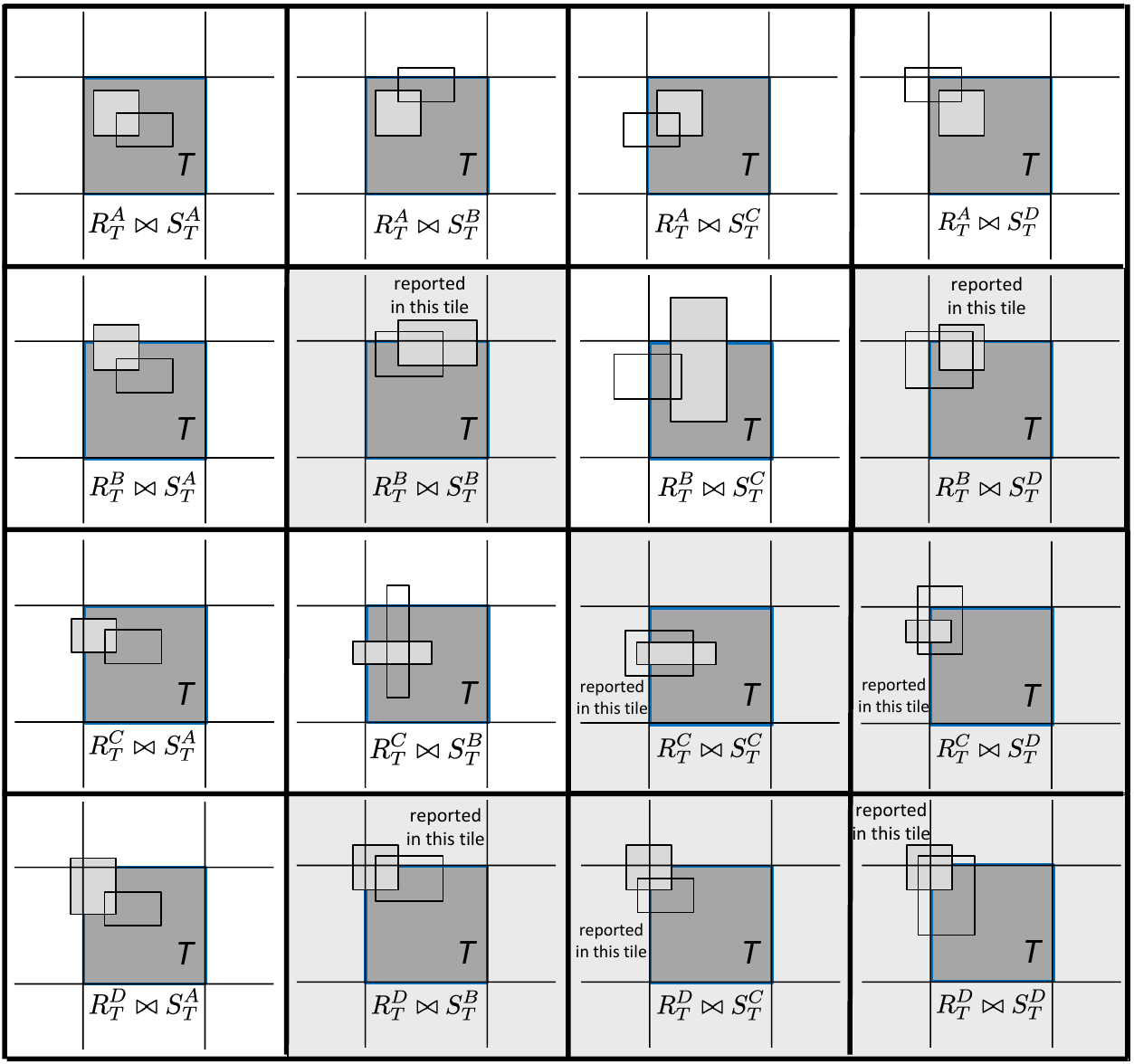}
  \caption{Decomposition of tile $T$'s join into 16 mini-joins.}
  \label{fig:minijoins}
\end{figure*}
}

Looking deeper into these 16 cases, we can easily show that 7 out of these mini-joins (the shaded
cases in the figure) produce only duplicate results, i.e., join
results that will also be reported in another tile. For example, if two
rectangles of class $B$ in tile $T$ intersect (i.e., the pair is contained in the result of
$R^B_T \bowtie S^B_T$),
then they will
definitely intersect also in another tile above $T$. 
We can also show that the remaining 9 mini-joins produce
only results that cannot be reported in any previous tile (in the $x$ or $y$
dimension or in both), but could be produced as duplicates in
some of the 7 shaded joins in a tile {\em after} $T$ in one or both
dimensions.
Hence, we 
\emph{never} evaluate 
the 7 shaded mini-joins
and only evaluate the 9 remaining without the need
of duplicate elimination.

\subsubsection{Optimizations}
\label{sec:join:mini-joins:optimizations}
Any spatial join algorithm can be utilized to evaluate the set of 9
mini-joins on each tile $T$; even a nested-loops approach. Following
previous work on spatial joins
\cite{TsitsigkosBMT19,PatelD96,ArgePRSV98}, we adopt a plane-sweep
join approach which is shown to perform significantly faster than
nested-loops; specifically we adopt the plane-sweep algorithm in
\cite{BrinkhoffKS93}. Algorithm~\ref{algo:ps} illustrates our
plane-sweep mini-join. To apply the plane-sweep approach, the contents
of each class are sorted on the sweeping dimension (Line~1). Without
loss of generality, we consider $x$ as the sweeping dimension for the
rest of this section, and so rectangles are sorted by $r.x_l$ and
$s.x_l$.
Sorting can take place also during the construction of the two-layer scheme, if an input dataset is partitioned online.
For every pair $(r,s)$ determined by the sweeping process (i.e., \eat{object }rectangles whose projections on $x$ intersect), we test in Lines~6 and 14, if the rectangles also intersect in the second dimension $y$; i.e., if $r.y_l \leq s.y_l \leq r.y_u$ or $s.y_l \leq r.y_l \leq s.y_u$. 
To boost the computation of mini-joins, we next discuss how we can save on the comparisons performed for $(r,s)$ pairs, capitalizing on our second layer of partitioning.
\eat{
\begin{algorithm}[t]
\LinesNumbered
\Input{collections of intervals $R$ and $S$}
\Output{\eat{set $J$ of }all intersecting \eat{interval }pairs $(r,s) \in R\times S$}
\BlankLine
\textbf{sort} $R$ and $S$ by \point{start} domain point\;
$r \leftarrow$ first interval in $R$\;
$s \leftarrow$ first interval in $S$\;
\While{$R$ and $S$ not depleted}
{
	\If{$r.\point{start} < s.\point{start}$} 
	{
		$s' \leftarrow s$\;
		\While{$s' \ne$ null \textbf{\emph{and}} $r.\point{end} \geq s'.\point{start}$}
		{
			\textbf{output} $(r,s')$\comm*{update result}
			$s' \leftarrow$ next interval in $S$\comm*{scan forward}
		}

		$r \leftarrow$ next interval in $R$\; 
	}
	\Else
	{
		$r' \leftarrow r$\;
		\While{$r' \ne$ null \textbf{\emph{and}} $s.\point{end} \geq r'.\point{start}$}
		{
			\textbf{output} $(r',s)$\comm*{update result}
			$r' \leftarrow$ next interval in $R$\comm*{scan forward}
		}

		$s \leftarrow$ next interval in $S$\; 
	}
}
\caption{Plane-sweep join}
\label{algo:ps}
\end{algorithm}
}
\begin{algorithm}[t]
\begin{algorithmic}[1]
\small
\Require classes of rectangles $R_T$ and $S_T$
\State\textbf{sort} $R_T$ and $S_T$ by $r.x_l$\Comment if not already sorted
\While{$R_T$ and $S_T$ not depleted}
	\If{$r.x_l < s.x_l$} 
		\State $s' \leftarrow s$
		\While{$s' \ne$ null \textbf{and} $r.x_u \geq s'.x_l$}
			\If{$r.y_l\!\leq\!s'.y_l\!\leq\!r.y_u$ \textbf{or} $s'.y_l\!\leq\!r.y_l\!\leq\!s'.y_u$}
				\State\textbf{output} $(r,s')$\Comment update result
			\EndIf
			\State$s' \leftarrow$ next rectangle in $S_T$\Comment scan forward
		\EndWhile
	\Else
		\State $r' \leftarrow r$
		\While{$r' \ne$ null \textbf{and} $s.x_u \geq r'.x_l$}
			\If{$r'.y_l\!\leq\!s.y_l\!\leq\!r'.y_u$ \!\textbf{or}\! $s.y_l\!\leq\!r'.y_l\!\leq\!s.y_u$}
				\State\textbf{output} $(r',s)$\Comment update result
			\EndIf
			\State$r' \leftarrow$ next rectangle in $R_T$\Comment scan forward
		\EndWhile
	\EndIf
\eat{
\rev{   \State $\mathcal{P}_T$ = sub-partitions of $T$ relevant to
   $W$ \Comment Lemmas \ref{lemma:filteringx} \& \ref{lemma:filteringy}
   \For{each sub-partition $T^X \in \mathcal{P}_T$}
      \State find all $r\in  T^X$ that intersect $W$ \Comment
      Sec. \ref{sec:compreduction}  \& \ref{sec:decomposition}
   \EndFor
}
}
\EndWhile
\end{algorithmic}
\caption{Plane-sweep mini-join}
\label{algo:ps}
\end{algorithm}

\stitle{Avoid unnecessary comparisons}. 
There exist two ways to utilize the $A$, $B$, $C$, $D$ classes in each tile $T$ for avoiding unnecessary rectangle comparisons. To understand the first, consider the $R_T^A \bowtie S_T^C$ mini-join. By definition, all objects rectangles in $S_T^C$ precede the rectangles in $R_T^A$. In practice, this means that we can apply a simplified version of plane-sweep for $R_T^A \bowtie S_T^C$ which performs forward scans on $R_T^A$ for each rectangle in $S_T^C$, as the $r.x_l > s.x_l$ holds by definition. This modification will save on unnecessary comparisons between $r \in R_T^A$ and $s \in S_T^C$ objects and further will allow us to avoid sorting the $S_T^C$ class. Algorithm~\ref{algo:reduced-ps} presents this reduced version of the plane-sweep mini-join. We can apply the same principle also for $R_T^A \bowtie S_T^C$, $R_T^C \bowtie S_T^A$, $R_T^A \bowtie S_T^D$, $R_T^D \bowtie S_T^A$, $R_T^B \bowtie S_T^C$ and $R_T^C \bowtie S_T^B$. Overall, we do not need to sort the $R_T^C$, $R_T^D$, $S_T^C$, $S_T^D$ classes.

\begin{algorithm}[t]
\begin{algorithmic}[1]
\small
\Require classes of rectangles $R_T$ and $S_T$
\State\textbf{sort} $R_T$ by $r.x_l$\Comment if not already sorted
\While{$S_T$ not depleted}
	\State $r' \leftarrow r$
	\While{$r' \ne$ null \textbf{and} $s.x_u \geq r'.x_l$}
			\If{$r'.y_l\!\leq\!s.y_l\!\leq\!r'.y_u$ \textbf{or} $s.y_l\!\leq\!r'.y_l\!\leq\!s.y_u$}
				\State\textbf{output} $(r',s)$\Comment update result
			\EndIf
			\State$r' \leftarrow$ next rectangle in $R_T$\Comment scan forward
	\EndWhile
\EndWhile
\end{algorithmic}
\caption{Reduced plane-sweep mini-join}
\label{algo:reduced-ps}
\end{algorithm}
Besides the plane-sweep process, we can also avoid unnecessary comparisons when considering the second dimension (i.e., $y$). The idea is similar to Lemmas~\ref{lemma:comp1} and \ref{lemma:comp2}\eat{ by replacing window $W$ with $S$ rectangles stored in the classes of tile $T$}. Take as example, the $R_T^A \bowtie S_T^B$ mini-join. For every pair $(r,s)$ of intersecting object rectangles in $x$, determined by plane-sweep, the $s.y_l < r.y_l$ condition holds by definition. Hence, to determine whether $r$, $s$ intersect also in dimension $y$ we only need to check the $r.y_l < s.y_u$ condition. The same principle can be used for the $R_T^A \bowtie S_T^D$, $R_T^B \bowtie S_T^A$, $R_T^D \bowtie S_T^A$, $R_T^B \bowtie S_T^C$ and $R_T^C \bowtie S_T^B$ mini-joins. The following lemma summarizes this optimization for every object rectangle $r$ in classes $R_T^B$, $R_T^D$ with $s$ in classes $S_T^A$ and $S_T^C$; the other case is symmetric.
\begin{lemma}
If an object rectangle $r$ in tile $T$ starts before the tile in the $y$ dimension, i.e., $T.y_l > r.y_l$, then $r$ intersects all $s$ rectangles in $T$ that start after $T.y_l$ with $r.y_u > s.y_l$.
\end{lemma}




\stitle{Avoid redundant comparisons}. 
\begin{figure}[t]
\centering
\begin{tabular}{cc}
\multicolumn{2}{c}{\includegraphics[width=0.7\columnwidth]{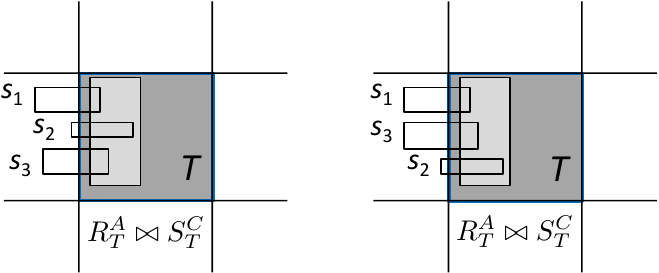}}\\
~~~~(a) $S_T^C$ unsorted &~~~~~~~~(b) $S_T^C$ sorted
\end{tabular}
\caption{Avoiding redundant comparisons on $R^A_T \bowtie S^C_T$; $R$ rectangle filled in light gray color.}
\label{fig:approach1}
\end{figure}
To illustrate this optimization, consider the $R^A_T \bowtie S^C_T$ mini-join and the \eat{object }rectangles in Figure~\ref{fig:approach1}(a). As already discussed, the reduced plane-sweep approach (Algorithm~\ref{algo:reduced-ps}) does not sort the $S^C_T$ class, to evaluate this mini-join; by definition, the start $s.{x_l}$ for their contents precede the $r.x_l$ start of all rectangles in $R^A_T$. Hence, assume that the $S^C_T$ rectangles are examined in the $s_1$, $s_2$, $s_3$ order. Rectangles $s_1$ and $r$ intersect in the $x$ dimension as $s_1.x_u > r.x_l$ holds and so, they are next compared in the $y$ dimension. Algorithm~\ref{algo:reduced-ps} will similarly determine that both $s_2$ and $s_3$ also intersect $r$ in $x$ by checking $s_2.x_u > r.x_l$ and $s_3.x_u > r.x_l$, respectively. However, since $s_2.x_u > s_1.x_u$ and $s_3.x_u > s_1.x_u$ hold, the $s_1.x_u > r.x_l$ check for $s_1$ automatically implies that $s_2.x_u > r.x_l$ and $s_3.x_u > r.x_l$ also hold. In other words, we conducted two extra comparisons to determine the intersecting $(r,s_2)$ and $(r,s_3)$ pairs. To avoid such redundant comparisons, we can sort $S^C_T$ by $s.x_u$ which will essentially allow us to determine intersecting rectangles in batches. Figure~\ref{fig:approach1}(b) illustrates this idea; all three intersecting pairs can be determined after checking only $s_1$ against $r$. Algorithm~\ref{algo:reduced-ps-grouping} modifies Algorithm~\ref{algo:reduced-ps} accordingly. After identifying the first rectangle $s$ that intersects current $r'$ in the $x$ dimension (Line~5), the algorithm pairs $r'$ with every rectangle $s'$ that follows $s$ in $S_T$ (Line~6).
\begin{algorithm}[t]
\begin{algorithmic}[1]
\small
\Require classes of rectangles $R_T$ and $S_T$
\State\textbf{sort} $R_T$ by $r.x_l$\Comment if not already sorted
\State\textbf{sort} $S_T$ by $r.x_u$\Comment if not already sorted
\While{$S_T$ not depleted}
	\State $r' \leftarrow r$
	\While{$r' \ne$ null \textbf{and} $s.x_u \geq r'.x_l$}
		\For{each rectangle $s'$ after $s$ in $S_T$}\Comment batch
			\If{$r'.y_l\!\leq\!s'.y_l\!\leq\!\!r'.y_u$ \!\textbf{or}\! $s'.y_l\!\leq\!r'.y_l\!\leq\!s'.y_u$}
				\State\textbf{output} $(r',s')$\Comment update result
			\EndIf
			\State$r' \leftarrow$ next rectangle in $R_T$\Comment scan forward
		\EndFor
	\EndWhile
\EndWhile
\end{algorithmic}
\caption{Reduced plane-sweep mini-join with batch outputting}
\label{algo:reduced-ps-grouping}
\end{algorithm}





\subsection{Join Strategies}
\label{sec:join:strategies}
We last discuss different join strategies depending on the (pre)-existence of two-layer partitioning in the input\eat{ datasets}. Following the classification in \cite{2011Mamoulis}, we consider three settings.

\eat{
\begin{figure*}[ht]
\centering
\begin{tabular}{cccc}
\includegraphics[width=0.23\textwidth]{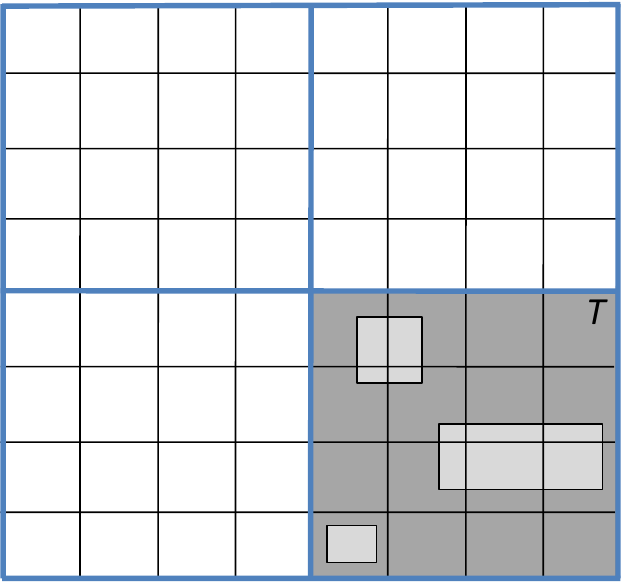}
&\includegraphics[width=0.23\textwidth]{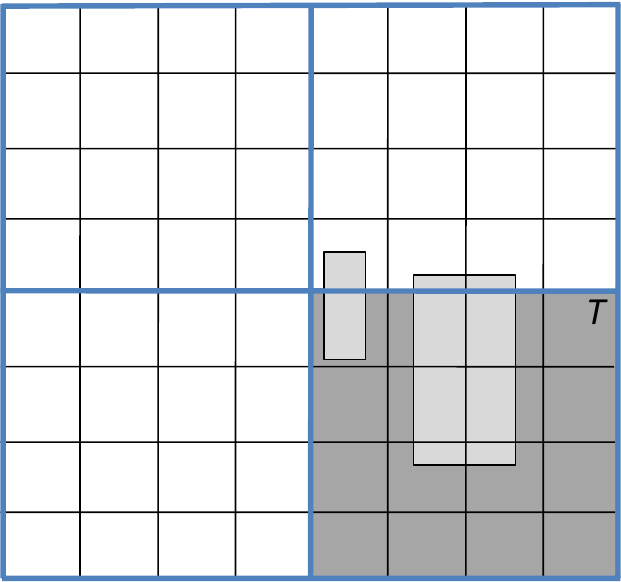}
&\includegraphics[width=0.23\textwidth]{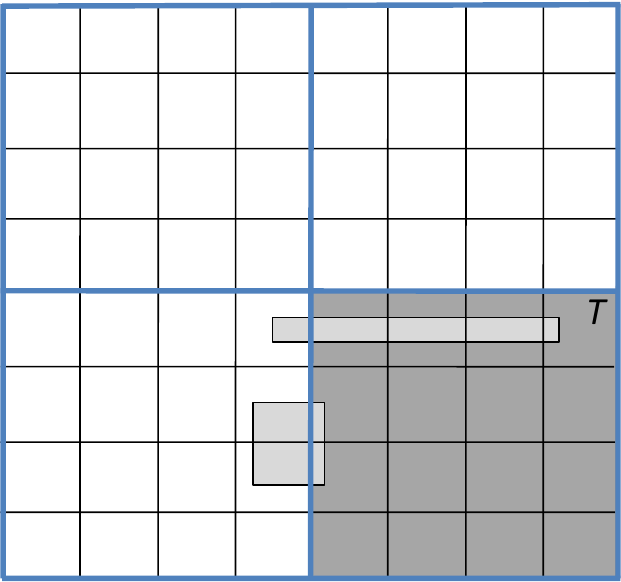}
&\includegraphics[width=0.23\textwidth]{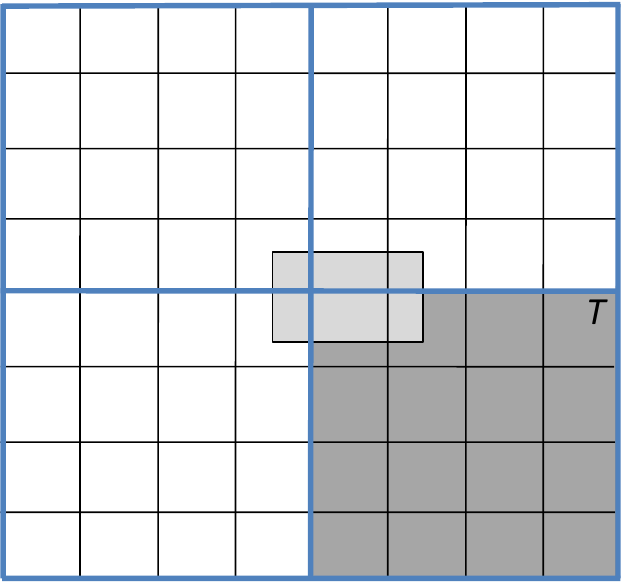}\\
objects of class A
&objects of class B
&objects of class C
&objects of class D
\end{tabular}
\caption{\todo{write me}}
\label{fig:trans}
\end{figure*}
}
\subsubsection{Both Inputs Indexed}
\label{sec:join:strategies:bothindexed}
Under this setting, a two-layer partitioning already exists on each input dataset $R$, $S$ to answer other types of spatial queries, e.g., \eat{window and disk }range queries. We distinguish between two cases with respect to the granularity of the pre-existing grids. If the two grids are identical, we can directly apply the mini-joins approach and its optimization\eat{ technique}s \eat{proposed }in Section~\ref{sec:join:mini-joins}. 

If the pre-existing grids have different granularities, then the
mini-joins approach is not directly applicable. A straightforward
solution \eat{to this problem }is to re-index one of the input
datasets, e.g. $R$, by creating a temporary two-layer partitioning
with a grid granularity that matches the grid on $S$,
similar to \cite{SabekM17}.
In this context, a key question that naturally arises is which input we should re-index. Typically, we could select the dataset with either the smallest cardinality or the smallest average object extent, because such a decision is expected to incur the lowest online indexing cost. We elaborate on this decision in our experimental analysis.

\begin{figure}[t]
\centering
\includegraphics[width=0.6\columnwidth]{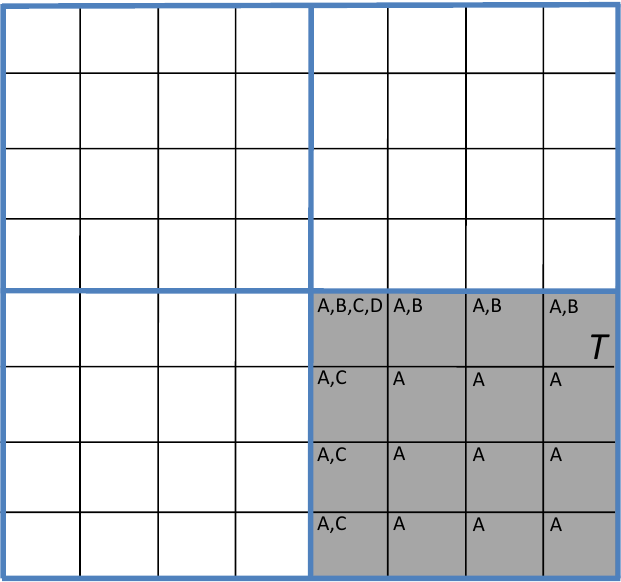}
\caption{Online grid transformation; $R$ partitioned by a $2\times 2$ grid, highlighted in blue, $S$ partitioned a $8\times 8$ grid, in black.}
\label{fig:trans}
\end{figure}
As an alternative, we also devise a new solution which completely eliminates the need to decide which input should be re-indexed and also significantly reduces the online indexing costs. For this purpose, the granularity of the pre-existing grids must be in a power of 2. If so, we can always define an online transformation of the finer grid to the coarser, by means of standard window range queries. Figure~\ref{fig:trans} exemplifies this transformation when $R$ is partitioned by a $2\times 2$ grid (colored in blue), and $S$ by a $8\times 8$ grid (in black). After overlaying the two grids, we observe that every tile $T$ in the $R$ grid overlaps a collection of exactly 16 adjacent tiles from the $S$ grid. Therefore, to re-partition $S$ to the $2\times 2$ grid of $R$, it suffices to compute the window queries defined by all tiles in the $R$ grid; the key idea is to determine the contents of the new $S^A_T$, $S^B_T$, $S^C_T$ and $S^D_T$ classes based on the results of every $T$ window query. For this purpose, we can utilize the original two-layer partitioning on $S$. Specifically, consider the shaded tile $T$ from the $R$ grid in Figure~\ref{fig:trans}. Similar to Figure~\ref{fig:ABCDexam}, we mark the relevant classes for each tile from the original $S$ grid that overlaps with the window query based on $T$, as discussed in Section~\ref{sec:range}. We can now construct $S^A_T$ by unifying the objects contained in the $A$ class of all overlapping tiles. For $S^B_T$, we consider only the contents of the $B$ class inside the top border overlapping tiles, while for $S^C_T$, the left border overlapping ones. Finally, $S^D_T$ is identical to the $D$ class inside the top left corner overlapping tile. The above process can be generalized for any two pre-existing grids of granularity $n\times n$ and $m\times m$, where $n > m$ and $n$, $m$ are powers of 2. Every tile in the second, coarser grid defines a window query that overlaps with exactly $(n/m)^2$ tiles from the first, finer one.

We developed two variants for the above transformation. The first constructs a temporary two-layer partitioning on input $S$ by matelializing the contents of each new  $S^A_T$, $S^B_T$, $S^C_T$ and $S^D_T$ classes. After this, we can directly utilize the mini-joins approach in Section~\ref{sec:join:mini-joins} to compute the $R\bowtie S$ spatial join. In contrast, the second variant never actually constructs this new two-layer scheme on input $S$. Instead, we adjust the joining process in Section~\ref{sec:join:mini-joins} to determine on-the-fly which tiles from the original grid on $S$ should be used in the mini-joins. Specifically, the $R^A_T \bowtie S^A_T$ is further decomposed to $(n/m)^2$ mini-joins, i.e., $R^A_T \bowtie S^A_T = \bigcup_i R^A_T \bowtie S^A_{T_i}$, where $T_i$ denotes one of the $(n/m)^2$ tiles from the original $S$ grid overlapping with tile $T$ from the $R$ grid. In the same spirit, $R^A_T \bowtie S^B_T$, $R^A_T \bowtie S^B_T$, $R^B_T \bowtie S^C_T$ and $R^C_T \bowtie S^B_T$ are decomposed into $(n/m)$ mini-joins, while $R^B_T \bowtie S^A_T$, $R^C_T \bowtie S^A_T$ and $R^D_T \bowtie S^A_T$ into $(n/m)^2$. 
In Section~\ref{sec:exps}, we compare the two variants and break down their total execution time.

\subsubsection{One Input Indexed}
\label{sec:join:strategies:oneindexed}
Under this setting, a two-layer partitioning already exists only for one of the input datasets; without loss of generality, assume for $S$. In this case, there exist two alternative approaches for computing the $R\bowtie S$ spatial join. The first is to construct a temporary two-layer partitioning on $R$ with a grid of identical granularity to the grid on $S$, such that we can directly apply the joining process in Section~\ref{sec:join:mini-joins}. Despite its simplicity, this approach may exhibit high total execution times because of the online indexing cost.

Alternatively, we can adopt an index-based join approach. According to this, we scan the contents of the $R$ input and probe the index on $S$. Specifically, we issue a window range query for each object rectangle $r$ in $R$ which is evaluated using the two-layer partitioning on $S$. 
To further enhance the performance of this approach, we can examine the objects in $R$ according to their position in space, instead in a random order, e.g., by first partitioning them online with a grid or using a space filling curve. This way, window queries for objects in $R$ that overlap neighboring parts of the $S$ grid are evaluated in nearby timestamps, improving the cache locality. 

\subsubsection{No Input Indexed}
\label{sec:join:strategies:noindex}
Under this setting, none of the input datasets $R$, $S$ is indexed\eat{ by our two-layer partitioning}. In this case, 
we can partition both inputs under identical grids to construct two temporary two-layer partitioning schemes and then directly apply the mini-joins approach from Section~\ref{sec:join:mini-joins}.

Since both inputs are indexed online, we can further enhance the join process by adopting a specialized storage optimization. Note that such an optimization cannot be utilized for the \eat{`both inputs indexed' or the 'one input indexed' }settings discussed in the previous sections, as at least one of the two-layer schemes used for the join, pre-exists in order to evaluate other types of queries. Essentially, if we enforce this optimization, the resulting two-layer partitioning will no longer be able to fulfil its original purpose.

Conventionally, each rectangle $r$ is stored as a quintuple $\langle id, r.x_l, r.x_u, r.y_l, r.y_u\rangle$. Assuming $x$ as the sweeping dimension (the other case is symmetric), $r.y_l$ is never needed for classes $B$ and $D$ when we check whether two rectangles intersect also in the $y$ dimension, i.e., in Lines~6 and 14 of Algorithm~\ref{algo:ps}, Line~5 of Algorithm~\ref{algo:reduced-ps} and in Line~7 of Algorithm~\ref{algo:reduced-ps-grouping}. This is because all contained rectangles start before the start of the tile in $y$. In fact, for class $D$, we do not need $r.x_l$ either, since $D$ is only joined to an $A$ class from the other input $S$ and its contents always precede those $S$ rectangles in both dimensions. Hence\eat{Under this premise}, a temporary two-layer partitioning stores all information for rectangles in classes $A$, $C$ but $\langle id, r.x_l, r.x_u, r.y_u\rangle$ for $B$ and $\langle id, r.x_u, r.y_u\rangle$ for $C$, which reduces the online partitioning costs.

\revisions{
\subsection{Extension to other SOPs}
\label{sec:join:sops}
The two-layer partitioning join in Section~\ref{sec:join:mini-joins} is directly applicable to any SOP, provided that identical partitions of the space (defined either offline or online) are considered for both inputs. Similarly, the index-based join approach can be applied when one of the inputs is indexed any SOP, enhanced with our two-layer partitioning. Lastly, when both inputs are indexed by the same SOP but under a different set of partitions, the re-partitioning approach in Section~\ref{sec:join:strategies:bothindexed} can be applied when there exists an alignment among the partitions; e.g., in quad-trees, for each a quadrant (of a coarse granularity) in one input that entirely covers a set of finer quadrants from the other.
}

}
\section{Experimental Evaluation}
\label{sec:exps}

Our analysis was conducted on a machine with 384 GBs of RAM and a dual
Intel(R) Xeon(R) CPU E5-2630 v4 clocked at 2.20GHz running CentOS Linux 7.6.1810.
All methods were implemented in C++, compiled using \texttt{gcc} (v4.8.5) with flags \texttt{-O3}, \texttt{-mavx} and \texttt{-march=native}.
\begin{table}[t]
\centering
\caption{Real-world datasets used in the experiments}
\label{tab:real}
\footnotesize
\eat{
\begin{tabular}{|@{~}c@{~}|@{~}c@{~}|@{~}c@{~}|@{~}c@{~}|@{~}c@{~}|}\hline
\textbf{dataset}		&\textbf{type}	&\textbf{card.}	&\textbf{avg. $x$-extent}	&\textbf{avg. $y$-extent}\\\hline\hline
ROADS					&linestrings		&$20$M					&$0.00001173$ 				&$0.00000915$ \\
EDGES					&polygons		&$70$M 					&$0.00000491$ 			&$0.00000383$ \\
\revisions{ZCTA5}					&polygons		&\todo{}	&\todo{}	&\todo{}\\
TIGER					&mixed				&$98$M					&$0.00000740$						&$0.00000576$\\
}
\begin{tabular}{|c|c|c|c|c|}\hline
\multirow{2}{*}{\textbf{dataset}}		&\multirow{2}{*}{\textbf{type}}	&\multirow{2}{*}{\textbf{cardinality}}			&\multicolumn{2}{c|}{\textbf{avg. relative [\%]}}\\\cline{4-5}
									&								&						&\textbf{$x$-extent}		&\textbf{$y$-extent}\\\hline\hline
ROADS					&linestrings		&$19$M					&$0.007$ 						&$0.013$ \\
EDGES					&polygons		&$69$M 					&$0.003$ 						&$0.005$ \\
\revisions{ZCTA5}					&polygons		&$33$K					&$1.7$							&$2.052$ \\
TIGER					&mixed			&$97$M					&$0.004$							&$0.008$\\
\hline
\end{tabular}
\end{table}

\stitle{Datasets.} We experimented with publicly
available Tiger 2015 datasets
\cite{EldawyM15},
\revisions{downloaded from
  \href{http://spatialhadoop.cs.umn.edu/datasets.html}{http://spatialhadoop.cs.umn.edu/datasets.html}. Their
  statistics are}
summarized in Table~\ref{tab:real}.
The fourth dataset resulted by merging all polygon and linestring
Tiger 2015 objects\eat{, excluding zip codes, counties and states}.
The objects in each dataset were normalized
so that the coordinates in each dimension take values inside $[0, 1]$.
The last two columns of the table are the relative (over the entire
space) average length for every object's MBR at each axis.
%
To test the robustness of our index, we also experimented
with synthetically generated datasets with rectangles of uniform and
zipfian spatial distribution.
Table \ref{tab:synthetic} shows the parameters used in data
generation.
The coordinates in each dimension take values in $[0, 1]$
and all generated rectangles in a dataset have the same area.
The width to height ratio of each rectangle was generated randomly in
the range $[0.25,4]$ to avoid unnaturally narrow rectangles.

\begin{table}[t]
\centering
\caption{Synthetic datasets (MBRs) used in the experiments}
\footnotesize
\label{tab:synthetic}
\begin{tabular}{|@{~}l@{~}|@{~}c@{~}|@{~}c@{~}|}\hline
\textbf{parameter}	&\textbf{values}	&\textbf{default}\\\hline\hline
cardinality					&1M, 5M, 10M, 50M, 100M	&10M\\
area				&\rev{$10^{-\infty}$}, $10^{-14}$, $10^{-12}$, $10^{-10}$, $10^{-8}$, $10^{-6}$	&$10^{-10}$\\
distribution				&Uniform or Zipfian ($a = 1$)			&---\\\hline
\end{tabular}
\end{table}

\stitle{Methods}. \ext{We implemented our secondary partitioning approach
denoted by \twolevel as part of a main-memory regular grid spatial index.
For each tile $T$ of the grid, \twolevel divides the (MBR, id)
pairs assigned to $T$ into four secondary partitions
($T^A$, $T^B$, $T^C$, $T^D$),
as discussed in Section \ref{sec:index}.}
%
%

\begin{table}[t]
\centering
\caption{Test methods and their throughput (range queries)\eat{ \todo{update me}}}
\footnotesize
\begin{tabular}{|@{~}c@{~}|@{~}l@{~}|@{~}c@{~}|c|c|}\hline
\multirow{2}{*}{\textbf{type}} &\multirow{2}{*}{\textbf{index}} &\multirow{2}{*}{\textbf{details}} &\multicolumn{2}{c|}{\textbf{throughput [queries/sec]}}\\\cline{4-5}
& & &ROADS &EDGES\\\hline\hline
\eat{
\multirow{4}{*}{SOP} 	&\twolevel								&Section~\ref{sec:index}										&24881		&7705\\
									&\twolevelplus						&Section~\ref{sec:decomposition}						&33090	&9994\\
									&\onelevel 							&grid with \cite{DittrichS00}									&12597 	&4403\\
}
\multirow{4}{*}{SOP} 	&\twolevel								&Section~\ref{sec:index}										&30981		&9406\\
									&\onelevel 							&grid with \cite{DittrichS00}									&12597 	&4403\\
\eat{
\multirow{4}{*}{SOP} 	&\twolevel								&Section~\ref{sec:index}										&30981		&9406\\
									&\twolevelplus						&Section~\ref{sec:decomposition}						&36444	&10855\\
									&\onelevel 							&grid using \cite{DittrichS00}								&13298 	&4618\\
}
									&\qtree 								&\cite{FinkelB74}	using \cite{DittrichS00}				&10949 	&3640\\
									&\rev{\qtree, \twolevel}			&\rev{\cite{FinkelB74} + Section~\ref{sec:index}}	&\rev{16883}	&\rev{5831}\\\hline
\multirow{4}{*}{DOP}		
									&\rtree						 			&\cite{LeuteneggerEL97} (boost.org)					&7888		&2011\\
									&\rstar	 								&\cite{BeckmannKSS90} (boost.org) 					&6415		&1610\\
									&\block									&\cite{OlmaTHA17}												&$<1$		&$<1$\\
									&\mxcif 								&\cite{Kedem82}													&8			&2\\\hline
\end{tabular}
\label{tab:competitors}
\end{table}
We considered both SOP and DOP competitors to our \twolevel\eat{ and \twolevelplus}, summarized in Table~\ref{tab:competitors}. First regarding SOPs, the \onelevel index is an in-memory grid with identical primary
partitioning as our \twolevel, but  uses the reference point
approach \cite{DittrichS00} to perform duplicate elimination.
Comparing \onelevel to \twolevel\eat{ and \twolevelplus} shows the benefit
of our secondary partitioning scheme and the techniques we propose in
Section \ref{sec:range} for
duplicate avoidance and minimization of comparisons. 
The second SOP competitor is a \qtree implementation, which assigns
each object MBR to all quadrants it intersects. 
As soon as the contents of a quadrant exceed a predefined
maximum {\em capacity} (set to 1000, after tuning),
the quadrant is split to four; the rectangles are then
re-distributed in the four generated children and {\em
  replicated} if they span the division borders.
To avoid
extensive splitting of \qtree nodes in the case of extremely skewed
data, a {\em maximum tree depth} (=12) is set.
The reference point
approach \cite{DittrichS00} is also used for duplicate elimination.
\rev{We also implemented a version of \qtree that uses our approach
  instead of \cite{DittrichS00}.}
Regarding DOPs, we used two implementations of in-memory R-trees
from the highly optimized Boost. Geometry library (boost.org)%
\footnote{Recent benchmarks \cite{LoskotW19} showed the superiority of
Boost.Geometry R-tree implementations over the ones in
libspatialindex.org}; an STR-bulkloaded \cite{LeuteneggerEL97} (denoted for simplicity as \rtree) and an \rstar
\cite{BeckmannKSS90}. 
Both trees have a fanout of 16 for inner and leaf
nodes; this configuration is reported to perform the best (we also
confirmed this by testing). 
The next DOP competitor is \block; the implementation was kindly provided
by the authors of \cite{OlmaTHA17}.
Finally, we also implemented and tested the \mxcif
for non-point data \cite{Kedem82}, which
stores each object at the lowest-level
quadrant which covers the object.
All compared methods are listed in Table \ref{tab:competitors}.

\stitle{Queries}. 
\eat{
We experimented with
both window and disk queries}
To study the evaluation of range queries, we experimented with window queries 
which
apply on non-empty areas of the map (i.e., they always return
results).
We vary  their relative area as a percentage of the
entire data space, inside the $\{0.01, 0.05, 0.1, 0.5, 1\}$ value range (default value $0.1$\% of the area of the
map).
Queries on synthetic data follow the same spatial distribution as the data.

\eat{
\subsection{Filtering vs. Refinement}
In the first experiment we evaluate the effectiveness of our
extra pre-refinement filtering (Lemma~\ref{lemma:ref}).
We used our \twolevel index
and considered three variants of query evaluation;
filtering is identical in all three variants.
Under \refvone, all candidates identified by the
filtering step are passed to the refinement step;
\refvtwo employs Lemma~\ref{lemma:ref}
to reduce the number of candidates to be refined;
last, \refvthree enhances \refvtwo by
using our secondary partitioning,
as discussed at the end of Section \ref{sec:refine}.
Figure~\ref{fig:ref} breaks down the average
execution time for 10000 window and disk queries;
note that for disk queries \refvthree is not applicable.
The pre-refinement filter
is very effective;
both \refvtwo and \refvthree significantly reduce the number
of candidates to be refined by over 90\%.
To achieve this however, they apply extra comparisons using the
MBRs;
these comparisons are more expensive
in the case of disk queries because they involve costly distance
computations between the disk center
and the corners of object MBRs.
When our secondary filtering technique is used,
the bottleneck of window queries is in the filtering step;
hence, in the subsequent experiments, we focus on the 
filtering step.

\begin{figure}[t]
\begin{center}
\begin{small}
\fbox{
{\scriptsize filtering}
\includegraphics[width=0.06\columnwidth]{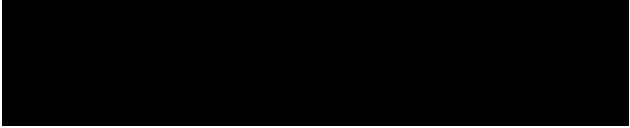}
\hspace{1ex}
{\scriptsize secondary filtering}
\includegraphics[width=0.06\columnwidth]{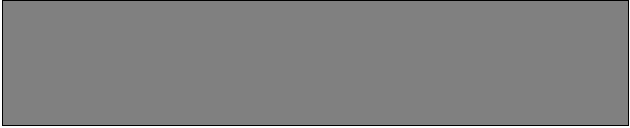}
\hspace{1ex}
{\scriptsize refinement}
\includegraphics[width=0.06\columnwidth]{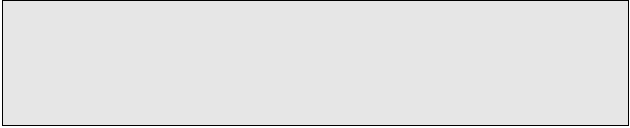}
\hspace{1ex}
}
\end{small}
\end{center}
\begin{tabular}{cc}
ROADS &EDGES\\
\hspace{-1ex}\includegraphics[width=0.485\columnwidth]{plots/grid_window_T8_ROADS_fixed_USA_c01_n10000_pBest_ref.pdf}
&\hspace{-1ex}\includegraphics[width=0.485\columnwidth]{plots/grid_window_T4_EDGES_fixed_USA_c01_n10000_pBest_ref.pdf}
\\
\multicolumn{2}{c}{window queries}\\
\hspace{-1ex}\includegraphics[width=0.485\columnwidth]{plots/grid_disc_T8_ROADS_fixed_USA_c01_n10000_pBest_ref.pdf}
&\hspace{-1ex}\includegraphics[width=0.485\columnwidth]{plots/grid_disc_T4_EDGES_fixed_USA_c01_n10000_pBest_ref.pdf}\\
\multicolumn{2}{c}{Disk queries}
\end{tabular}
\caption{Time breakdown in two-layer indexing \todo{update me}}
\label{fig:ref}
\end{figure}
}

\subsection{Indexing and Tuning}
\label{sec:exps:tuning}
\begin{figure}[t]
\eat{
\begin{center}
\begin{small}
\fbox{
{\scriptsize \onelevel}
\includegraphics[width=0.06\columnwidth]{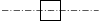}
\hspace{1ex}
{\scriptsize \twolevel}
\includegraphics[width=0.06\columnwidth]{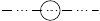}
}
\end{small}
\end{center}
}
\begin{tabular}{cc}
ROADS &EDGES\\
\hspace{-2ex}\includegraphics[width=0.485\columnwidth]{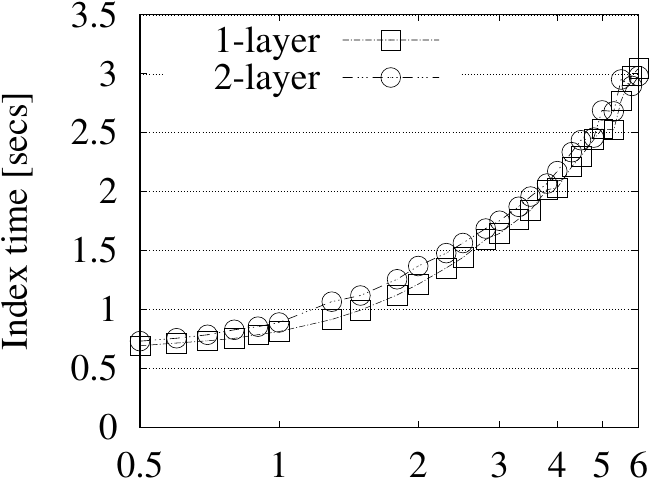}
&\hspace{-1ex}\includegraphics[width=0.485\columnwidth]{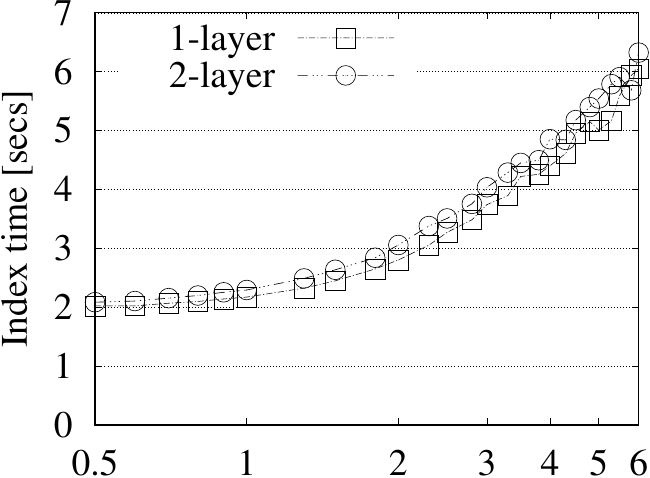}\\
\hspace*{-3ex}{\scriptsize partitions per dimension [$\times$1000]} &\hspace*{-2ex}{\scriptsize partitions per dimension [$\times$1000]}\\\\
\eat{
\hspace{-2ex}\includegraphics[width=0.485\columnwidth]{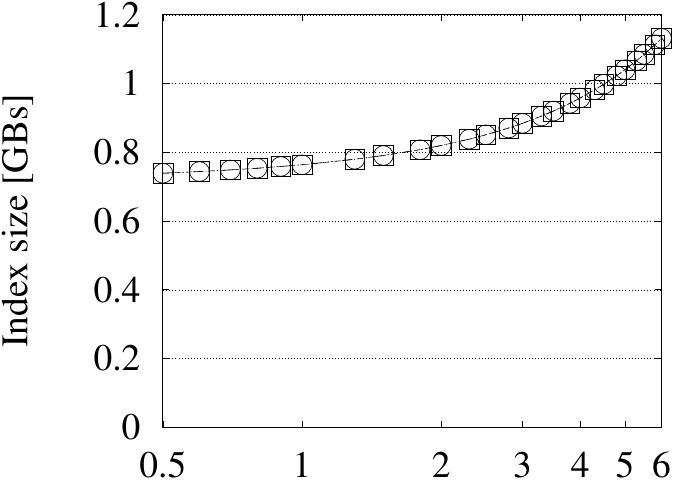}
&\hspace{-1ex}\includegraphics[width=0.485\columnwidth]{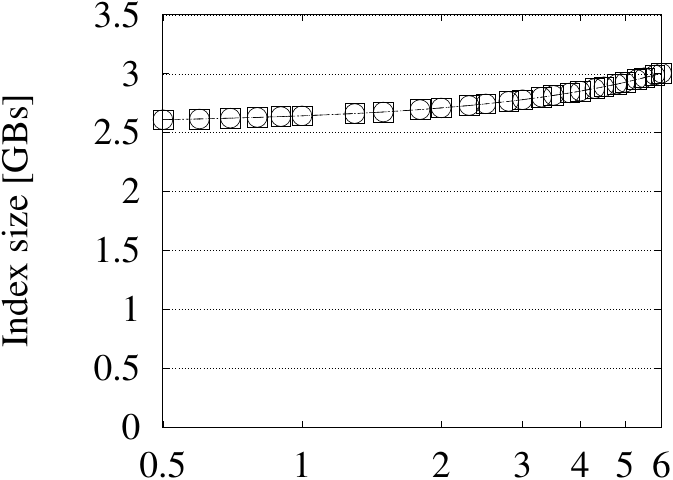}\\
\hspace*{-3ex}{\scriptsize partitions per dimension [$\times$1000]} &\hspace*{-2ex}{\scriptsize partitions per dimension [$\times$1000]}\\\\
}
%
\hspace*{-2ex}\includegraphics[width=0.485\linewidth]{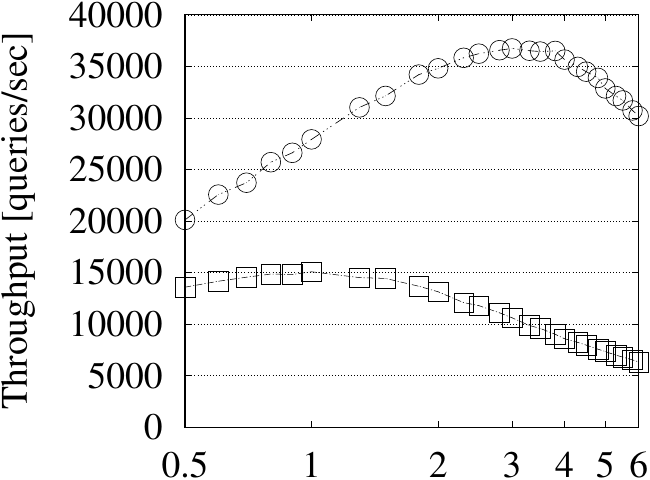}
&
\hspace*{-1ex}\includegraphics[width=0.485\linewidth]{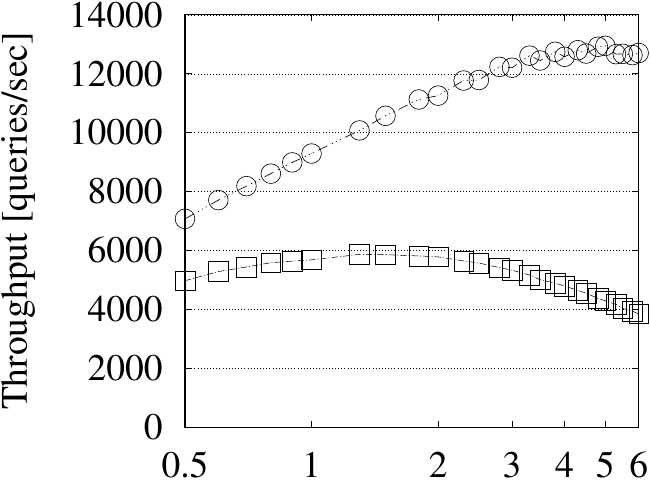}\\
\hspace*{-3ex}{\scriptsize partitions per dimension [$\times$1000]} &\hspace*{-2ex}{\scriptsize partitions per dimension [$\times$1000]}
\end{tabular}
\caption{Building \& tuning grid-based indices (range queries)}
\label{fig:indexcost}
\end{figure}

We first investigate the index building cost and  tuning.
\ext{The first two plots of Figure~\ref{fig:indexcost} compare the indexing times of 
\onelevel and \twolevel
on ROADS and EDGES\eat{ datasets}, while varying the granularity of the grid partitioning. Indexing sizes are omitted due to lack of space. Both indices occupy the same space, regardless of employing the secondary partitioning or not, as exactly the same number of object MBRs (originals
and replicas) are stored.
Note that the index sizes do not grow too much with the grid granularity, from 0.8 to 1.2GBs for ROADS, from 2.5 to 3GBs for EDGES, which means that MBR replication is not excessive.
}
Naturally, the indexing cost for both\eat{all three} indices rises as we
increase the granularity of the grid.
%
\eat{As expected, \onelevel and \twolevel have the same space
requirements;
regardless of employing secondary partitioning or not,
both indices store exactly the same number of object MBRs (originals
and replicas).
}
In terms of indexing time, \twolevel is
only slightly
more expensive
than \onelevel.
%
The construction costs for the two quadtrees (not shown) are
7s and 28.2s, respectively, and their sizes are similar to those of the
corresponding \onelevel indices.
The sizes of the packed R-trees (not shown) are about the same as the
sizes of the corresponding \onelevel, \twolevel indices,
indicating that the replication
ratio of our index is low.
In addition, the bulk loading costs of the R-trees are
5.2s and 19.5s for the two datasets, respectively,
\ext{both higher than the construction cost of \twolevel.}


The last two plots of Figure \ref{fig:indexcost}
compare  the window query throughputs of
\onelevel and \twolevel 
for different grid granularities.
The two\eat{three} methods achieve their best
throughputs when several thousands of partitions per dimension are used. 
Observe that
employing our secondary partitioning significantly
enhances query processing; \twolevel \eat{and \twolevelplus }always
outperforms \onelevel by a wide margin (2x--3x).
It is worth noting that \onelevel uses the comparisons reduction
optimization described in Section \ref{sec:compreduction}, meaning
that the performance gap is due to our
secondary partitioning\eat{ and the storage decomposition (by \twolevelplus)}.
Specifically, our approach outperforms
the state-of-the-art reference point method
for result deduplication \cite{DittrichS00} used in \onelevel
by a factor of at least 2.
For a wide range of granularities 
(i.e., 2000 to 5000 partitions per dimension), 
the throughput of both \eat{three }methods does not change
significantly meaning that finding the best granularity is not crucial
to query performance.
We observed similar trends on the TIGER and
on the synthetic datasets (not shown due to lack of space).
\revisions{As a rule of the thumb, we can set the extent of a grid cell in each dimension, 10 times larger than the average extent of the objects in that dimension.}
For the rest of our analysis, we used the best granularity for 
\onelevel and \twolevel, \revisions{see Table~\ref{tab:granularities}}.

\begin{table}
\centering
\caption{\revisions{Best granularities (partitions per dimension); in parenthesis, the power of 2 used for the transformation-based methods in Figure~12.}}
\label{tab:granularities}
\begin{tabular}{|@{~}c@{~}|@{~}c@{~}|@{~}c@{~}|@{~}c@{~}|@{~}c@{~}|@{~}c@{~}|@{~}c@{~}|}
\hline
\revisions{index}		&\revisions{ROADS}					&\revisions{EDGES}					&\revisions{ZTCA5}			&\revisions{TIGER}					&\revisions{Uniform}		&\revisions{Zipfian}\\
\hline\hline
\revisions{\onelevel}		&\revisions{1000\eat{ ($2^{10}$)}}	&\revisions{3000\eat{ ($2^{11}$)}}	&\revisions{not used}			&\revisions{1000\eat{ ($2^{10}$)}}	&\revisions{500}			&\revisions{3000}\\
\revisions{\twolevel}		&\revisions{3000 ($2^{11}$)}			&\revisions{5000 ($2^{12}$)}			&\revisions{400 ($2^8$)}		&\revisions{5300\eat{ ($2^{12}$)}}	&\revisions{450}			&\revisions{3000}\\
\hline
\end{tabular}
\end{table}

\eat{
\begin{figure*}
\begin{center}
\begin{small}
\fbox{
{\footnotesize \rtree}
\includegraphics[width=0.06\columnwidth]{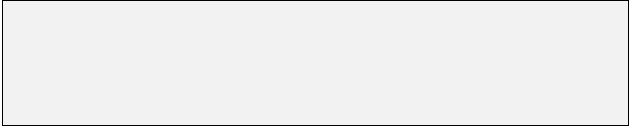}
\hspace{0.5ex}
{\footnotesize \qtree}
\includegraphics[width=0.06\columnwidth]{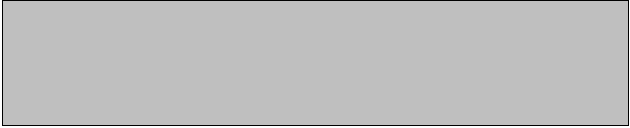}
\hspace{0.5ex}
{\footnotesize \onelevel}
\includegraphics[width=0.06\columnwidth]{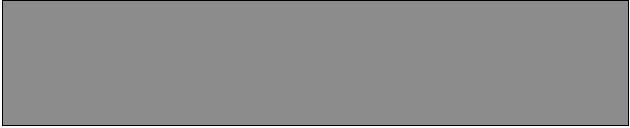}
\hspace{0.5ex}
{\footnotesize \twolevel}
\includegraphics[width=0.06\columnwidth]{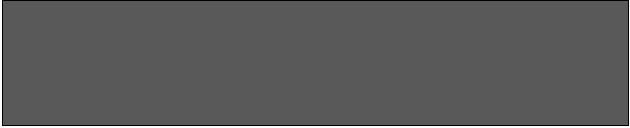}
\hspace{0.5ex}
{\footnotesize \twolevelplus}
\includegraphics[width=0.06\columnwidth]{figures/1.pdf}
}
\end{small}
\end{center}
\begin{tabular}{cccc}
\multicolumn{4}{c}{ROADS}\\
\hspace{-1ex}\includegraphics[width=0.485\columnwidth]{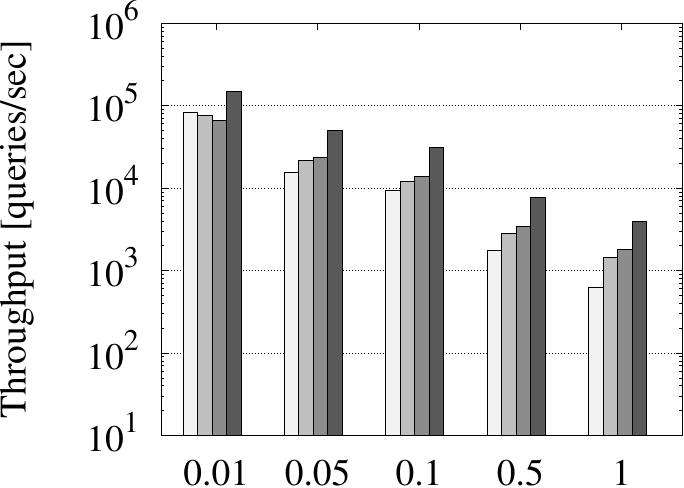}
&\hspace{-1ex}\includegraphics[width=0.485\columnwidth]{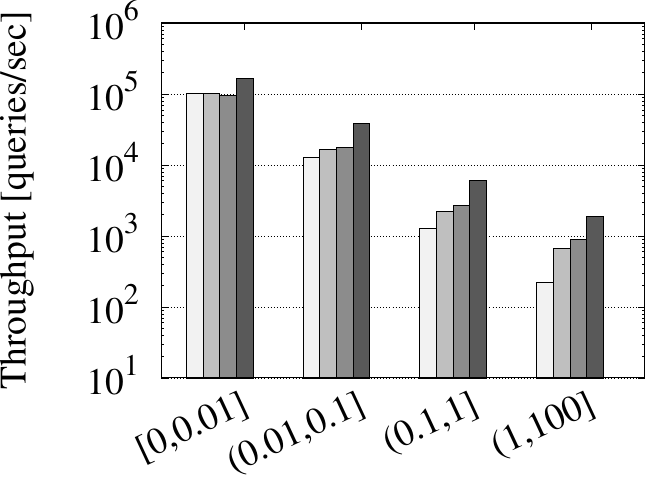}
&\hspace{-1ex}\includegraphics[width=0.485\columnwidth]{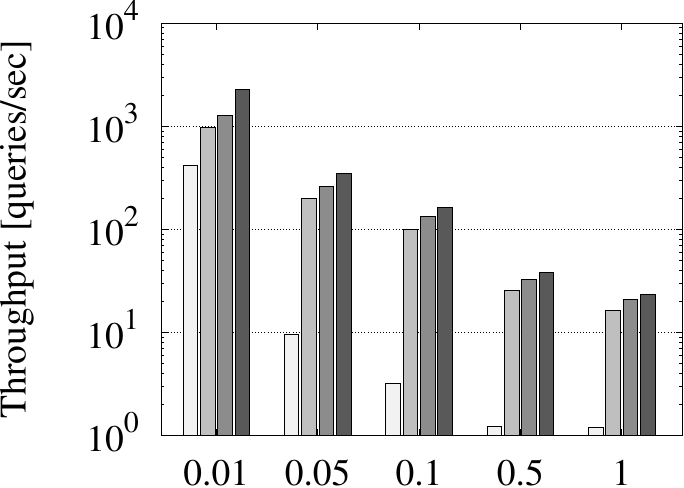}
&\hspace{-1ex}\includegraphics[width=0.485\columnwidth]{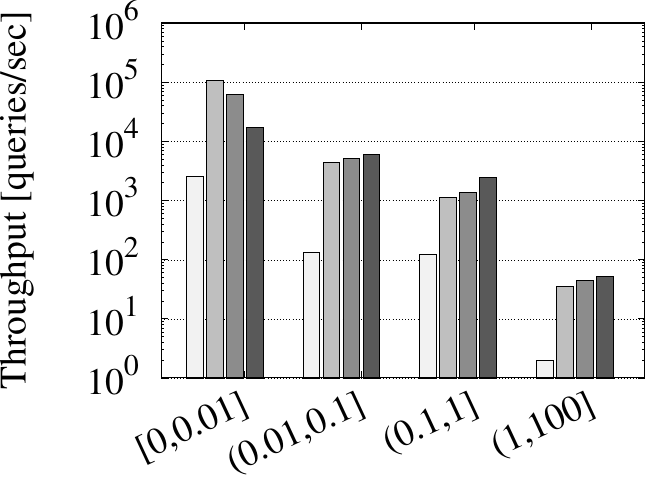}\\
{\scriptsize query relative area [\%]} &{\scriptsize query selectivity [\%]} &{\scriptsize query relative area [\%]} &{\scriptsize query selectivity [\%]}\\
\multicolumn{2}{c}{Window queries} &\multicolumn{2}{c}{Disk queries}\\\\
\multicolumn{4}{c}{EDGES}\\
\hspace{-1ex}\includegraphics[width=0.485\columnwidth]{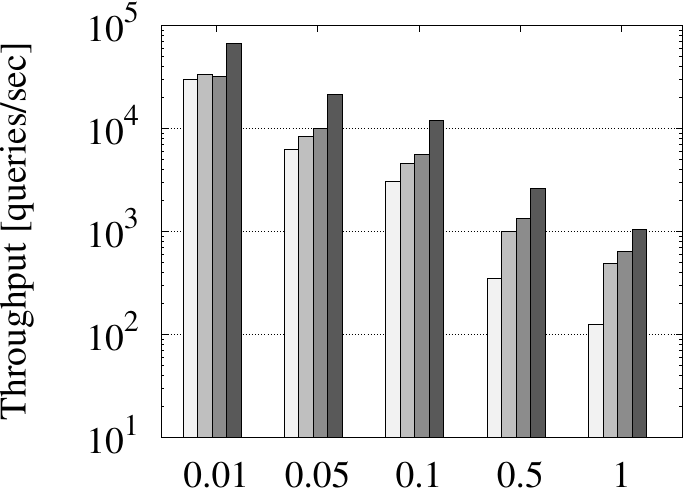}
&\hspace{-1ex}\includegraphics[width=0.485\columnwidth]{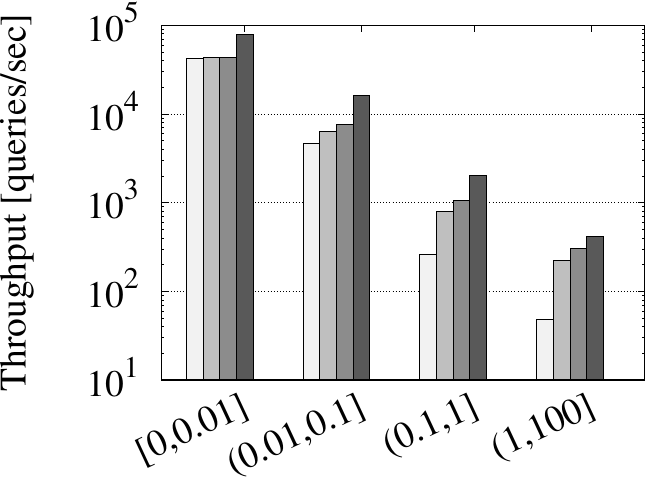}
&\hspace{-1ex}\includegraphics[width=0.485\columnwidth]{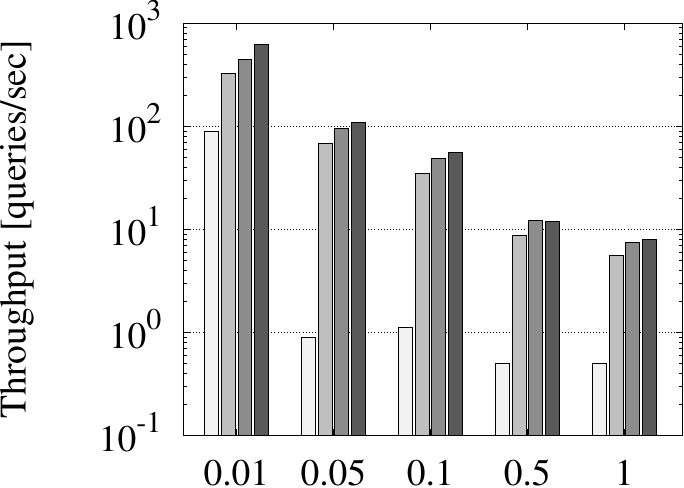}
&\hspace{-1ex}\includegraphics[width=0.485\columnwidth]{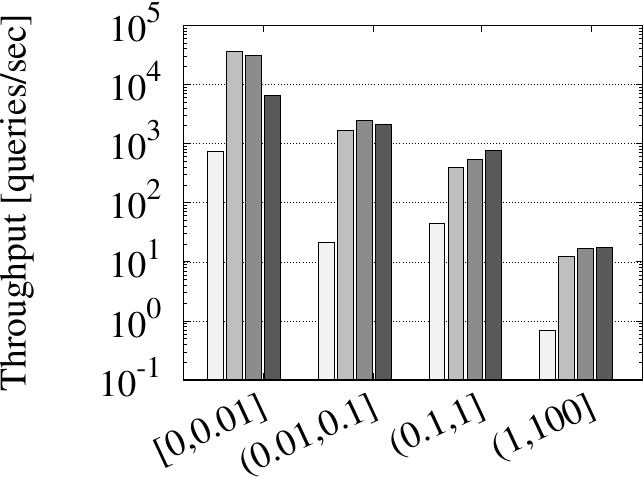}\\
{\scriptsize query relative area [\%]} &{\scriptsize query selectivity [\%]} &{\scriptsize query relative area [\%]} &{\scriptsize query selectivity [\%]}\\
\multicolumn{2}{c}{Window queries} &\multicolumn{2}{c}{Disk queries}\\\\
\multicolumn{4}{c}{TIGER}\\
\hspace{-1ex}\includegraphics[width=0.485\columnwidth]{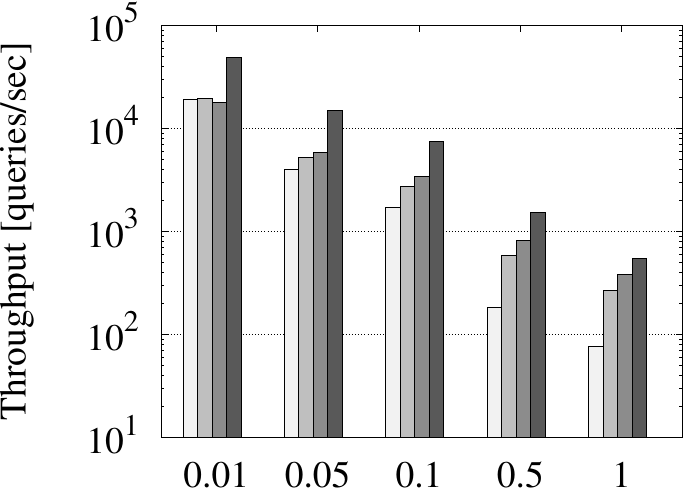}
&\hspace{-1ex}\includegraphics[width=0.485\columnwidth]{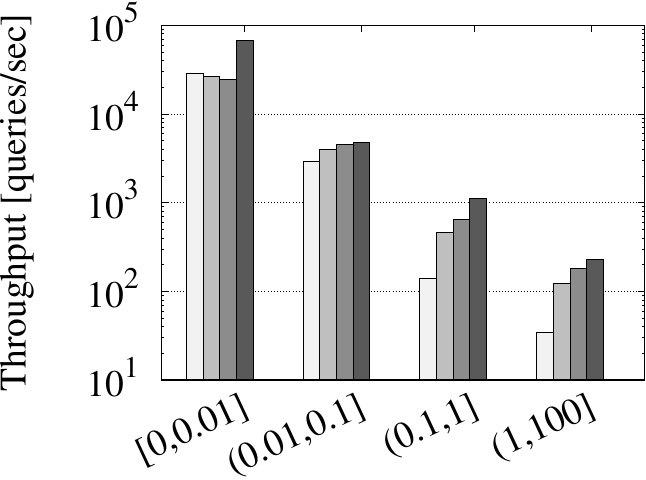}
&\hspace{-1ex}\includegraphics[width=0.485\columnwidth]{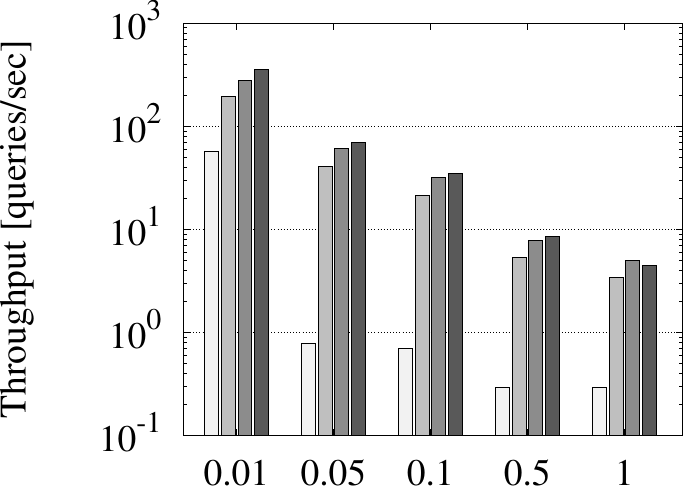}
&\hspace{-1ex}\includegraphics[width=0.485\columnwidth]{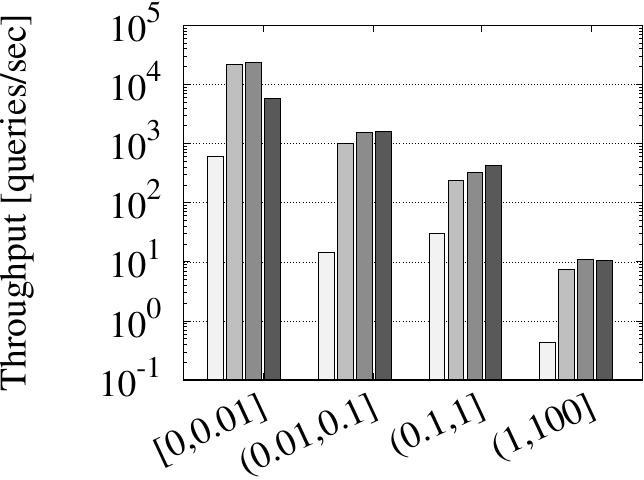}\\
{\scriptsize query relative area [\%]} &{\scriptsize query selectivity [\%]} &{\scriptsize query relative area [\%]} &{\scriptsize query selectivity [\%]}\\
\multicolumn{2}{c}{Window queries} &\multicolumn{2}{c}{Disk queries}
\end{tabular}
\caption{Query processing: real data \todo{update me}}
\label{fig:vary-c}
\end{figure*}
}
\begin{figure}
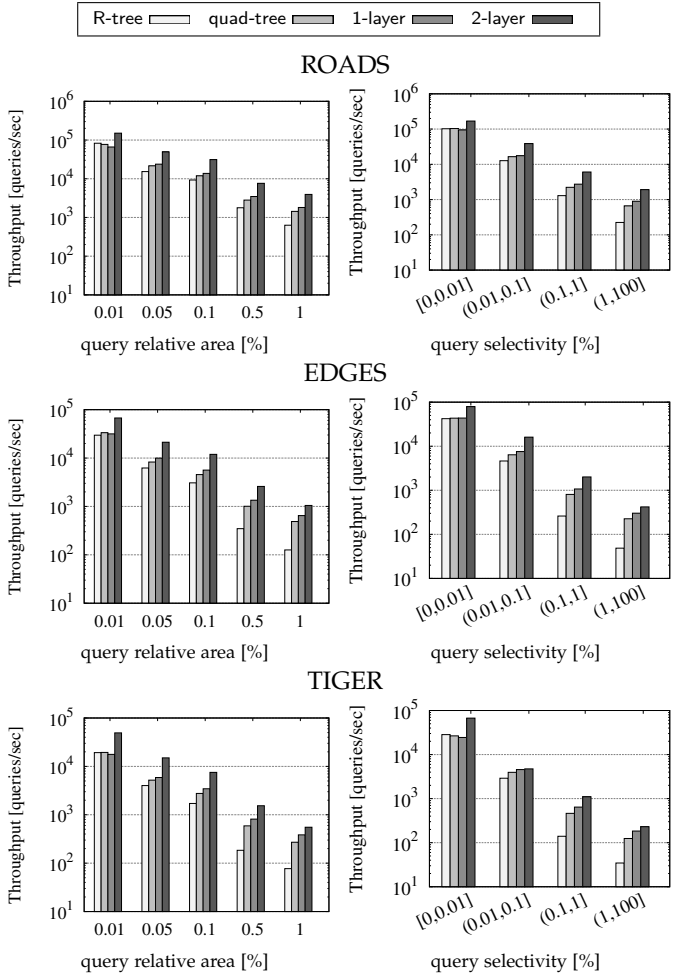

\begin{center}
\begin{small}
\fbox{
{\scriptsize \rtree}
\includegraphics[width=0.06\columnwidth]{figures/0_05.pdf}
\hspace{0.5ex}
{\scriptsize \qtree}
\includegraphics[width=0.06\columnwidth]{figures/0_25.pdf}
\hspace{0.5ex}
{\scriptsize \onelevel}
\includegraphics[width=0.06\columnwidth]{figures/0_45.pdf}
\hspace{0.5ex}
{\scriptsize \twolevel}
\includegraphics[width=0.06\columnwidth]{figures/0_65.pdf}
\hspace{0.5ex}
}
\end{small}
\end{center}
\begin{tabular}{cc}
\multicolumn{2}{c}{ROADS}\\
\hspace{-1ex}\includegraphics[width=0.48\columnwidth]{plots_new/window_throughput_T8_ROADS_fixed_mbr_USA_n10000_pBest_vary-c_comps_boxes.pdf}
&\hspace{-1ex}\includegraphics[width=0.48\columnwidth]{plots_new/window_throughput_T8_ROADS_fixed_mbr_USA_n10000_pBest_vary-s_comps_4_boxes.pdf}\\
{\scriptsize query relative area [\%]} &{\scriptsize query selectivity [\%]}\\
%
\multicolumn{2}{c}{EDGES}\\
\hspace{-1ex}\includegraphics[width=0.48\columnwidth]{plots_new/window_throughput_T4_EDGES_fixed_mbr_USA_n10000_pBest_vary-c_comps_boxes.pdf}
&\hspace{-1ex}\includegraphics[width=0.48\columnwidth]{plots_new/window_throughput_T4_EDGES_fixed_mbr_USA_n10000_pBest_vary-s_comps_4_boxes.pdf}\\
{\scriptsize query relative area [\%]} &{\scriptsize query selectivity [\%]}\\
%
\multicolumn{2}{c}{TIGER}\\
\hspace{-1ex}\includegraphics[width=0.48\columnwidth]{plots_new/window_throughput_T_Important_fixed_mbr_USA_n10000_pBest_vary-c_comps_boxes.pdf}
&\hspace{-1ex}\includegraphics[width=0.48\columnwidth]{plots_new/window_throughput_T_Important_fixed_mbr_USA_n10000_pBest_vary-s_comps_4_boxes.pdf}\\
{\scriptsize query relative area [\%]} &{\scriptsize query selectivity [\%]}\\
\end{tabular}
\caption{Range queries: real datasets}
\label{fig:vary-c}
\end{figure}
\subsection{Range Query Performance}
\ext{We next compare our two-layer partitioning against competition on\eat{in terms of} query throughput.}\eat{ (window
and disk queries),
evaluate batch and parallel query processing, and finally measure
their update costs.

\stitle{Window queries}.
First, we report in
}
Table \ref{tab:competitors} reports the throughput (queries/sec)
achieved by each index for 10K window queries (of average relative area
0.1\% of the map) on ROADS and EDGES.
\twolevel outperforms the competition by a wide
margin.
\rev{Note that the performance of
  \qtree is greatly improved by our secondary partitioning, which confirms
  that other SOPs besides grids can benefit from our approach.
  However, as \twolevel is consistently more
  efficient than the \qtree using our approach,
  we do not consider the latter in the rest of the experiments.}
\rtree is the most efficient DOP competitor, outperforming \rstar
(from the same library).
\block takes seconds to evaluate range queries on our datasets, which can be attributed to the fact that it is implemented for 3D objects. Similar, \mxcif is orders of magnitude slower than the \rtree.
Under these, in the rest of the tests we only include
\onelevel, \rtree and \qtree indices as the key competitors to our 
\twolevel.

\begin{figure}[t]
\begin{center}
\begin{small}
\fbox{
{\scriptsize \rtree}
\includegraphics[width=0.06\columnwidth]{figures/0_05.pdf}
\hspace{0.5ex}
{\scriptsize \qtree}
\includegraphics[width=0.06\columnwidth]{figures/0_25.pdf}
\hspace{0.5ex}
{\scriptsize \onelevel}
\includegraphics[width=0.06\columnwidth]{figures/0_45.pdf}
\hspace{0.5ex}
{\scriptsize \twolevel}
\includegraphics[width=0.06\columnwidth]{figures/0_65.pdf}
}
\end{small}
\end{center}
\centering
\begin{tabular}{cc}
Uniform (default) &Zipfian (default)\\
%
%
\hspace{-1ex}\includegraphics[width=0.48\columnwidth]{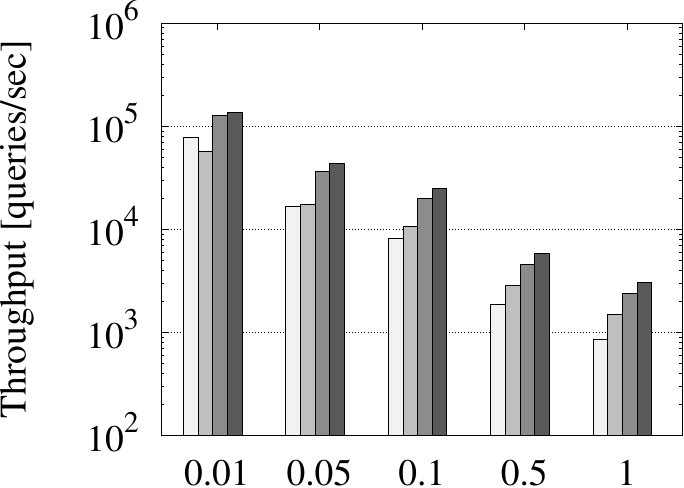}
&\hspace{-1ex}\includegraphics[width=0.48\columnwidth]{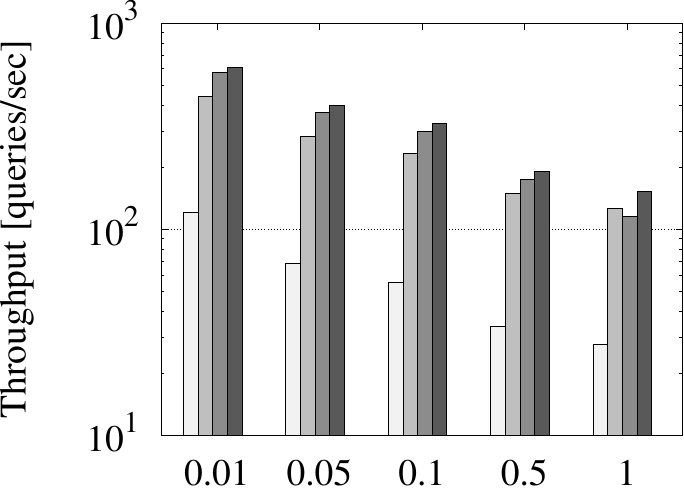}
\\
 {\scriptsize query relative extent [\%]
  } &{\scriptsize query relative extent [\%] }\\\\
%
%
\hspace{-1ex}\includegraphics[width=0.48\linewidth]{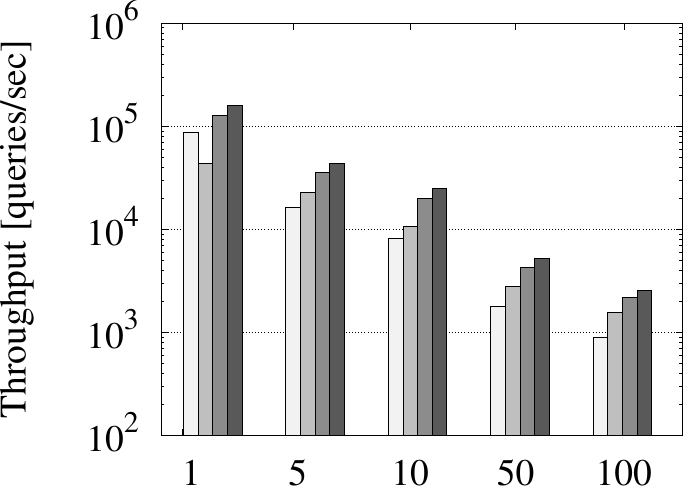}
&
\hspace*{-1ex}\includegraphics[width=0.48\linewidth]{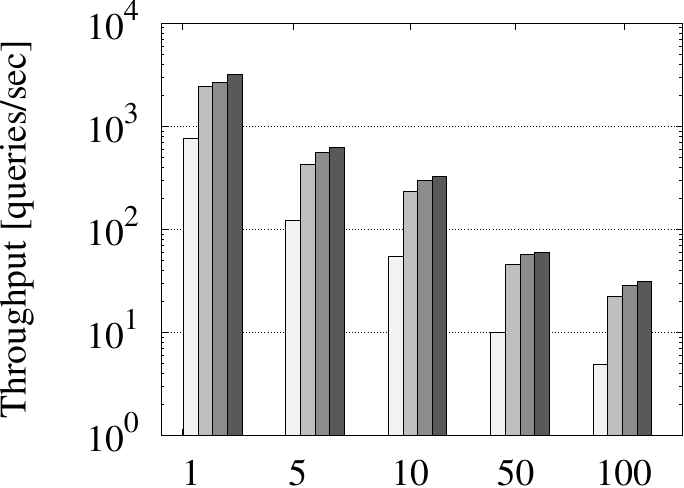}\vspace*{-1ex}\\
{\scriptsize data cardinality [$\times 10^6$]} &{\scriptsize data cardinality [$\times 10^6$] }\\\\
\hspace{-1ex}\includegraphics[width=0.48\linewidth]{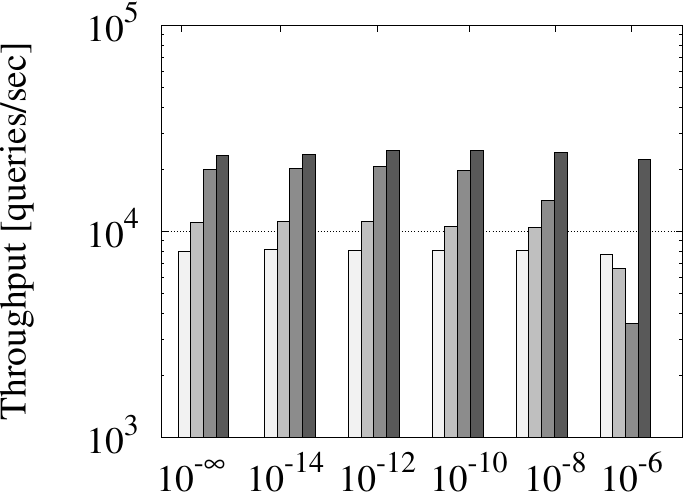}
&
\hspace*{-1ex}\includegraphics[width=0.48\linewidth]{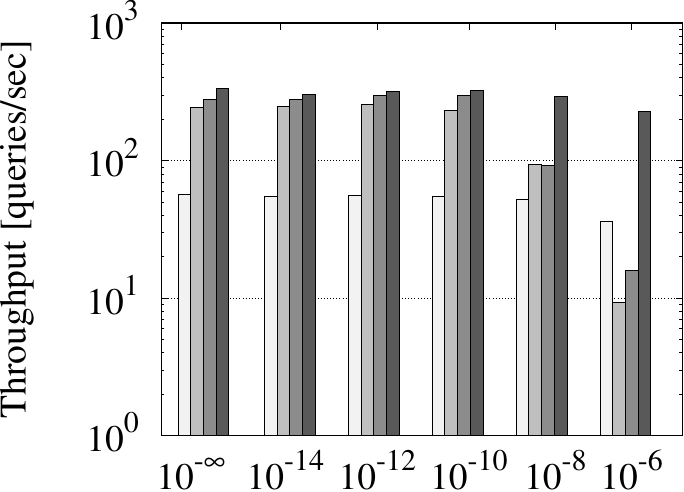}\vspace*{-1ex}\\
{\scriptsize \rev{data rectangle area}} &{\scriptsize \rev{data rectangle area}}
\end{tabular}
\caption{Range queries: synthetic datasets} 
\label{fig:synth}
\end{figure}

 The first two columns of Figure~\ref{fig:vary-c} show
the throughput of the four\eat{five} competitors
for window queries of varying relative area and selectivity on the three real datasets.
For the experiments of the second column, we collected the runtimes of
all queries (regardless of their areas) and averaged them after
grouping them by selectivity.
Naturally, query processing is negatively affected by both factors.
We also observe that \twolevel 
is consistently much faster than the competition on all datasets and
query areas\eat{; in addition, the relative difference between 
\twolevel and
\twolevelplus is stable}. 
For each query, \twolevel 
accesses the relevant partitions very fast (without the need of
traversing a hierarchical index) and
manage to drastically reduce the required number of computations.
Figure \ref{fig:synth} compares all methods for window queries on the
synthetic datasets \revisions{(for grid granularities, see Table~\ref{tab:granularities})}. In these experiments, we additionally vary the database
size and the areas of the
data objects.
In terms of query throughput w.r.t. query area and selectivity, the trends are similar
as those for the real data.
In addition, the data cardinality does not affect the
relative performance of the methods.
Finally, we observe that \twolevel 
is more robust
to the area of the data objects compared to the competition. As the area
grows, the replication to tiles increases, and so \onelevel and 
\qtree
have to compute
and eliminate more duplicate results. In contrast, \twolevel,
with the help of our secondary partitioning,
completely avoids the generation and elimination of duplicate results.
\rev{
  On the other hand when the data area shrinks
  ($10^{-\infty}$ represents the case of extremely small
  rectangles that resemble points), the replication to
tiles decreases.
\onelevel and the \qtree still need to perform the extra comparison of
the de-duplication reference test, which explains 
the stable advantage of \twolevel\eat{ and \twolevelplus}.}

\eat{
\stitle{Disk range queries}.
For disk range queries,
we report results on the real data
in the last two columns of
plots in Figure~\ref{fig:vary-c}.
\twolevelplus is not included in the comparison, because 
storage decomposition cannot improve distance computations.
For disk queries on \onelevel and \qtree, we cannot use the
reference point technique to
eliminate duplicate results
(and duplicate elimination using hashing is too expensive).
Thus, we implemented disk queries on them as
follows. We executed a window query using the MBR of the query range and
eliminated any duplicates intersecting the window. For all tiles/quadrants
inside the disk range, we just reported all window query results there as disk query
results. For all other tiles and quadrants we performed distance tests
before confirming and reporting the results.
The plots show once again the superiority of the \twolevel index.
}


\eat{
\stitle{Batch and Parallel Query Processing}.
\panos{Open issues: (1) we keep these exps if we of course keep Section~\ref{sec:batch}. (2) If we have exps for parallel joins as well, I believe there should be a separate subsection only for parallel processing; the non-parallel exps can stay here.}
Figure \ref{fig:vary-c_batch} compares the two approaches (\qatomic
and \tatomic), discussed in Section \ref{sec:batch}, for batch window query processing
(10K queries or 1\% relative area, per batch) on ROADS and EDGES.%
\footnote{Similar findings are observed for TIGER, but the results are omitted due to lack of space.}
A general observation from the plots is that \tatomic is superior to 
\qatomic when the dataset is large (i.e., dense) and the queries are
relatively large. In this case, the sizes of the dedicated tables for
each class per tile are large and cache conscious  \tatomic approach
makes a difference. On the other hand, the overhead of finding and accumulating
the subtasks per tile does not pay off when the number of queries
on each tile is too small or when the tiles do not contain many rectangles.
The advantage of \tatomic becomes more prominent in parallel query
processing.  Figure \ref{fig:vary-c_batch_parallel}
shows the speedup of batch query evaluation
on the two largest datasets
(again, 10K queries per
batch)
as a function of the
number of parallel threads.
Note that  \tatomic scales gracefully with the number of threads (up
to about 25 threads, where it starts being affected by
hyperthreading).
On the other hand, \qatomic scales poorly due to the numerous cache
misses.

\begin{figure}
\centering
\begin{tabular}{cc}
ROADS &EDGES\\
\hspace{-1ex}\includegraphics[width=0.485\columnwidth]{../plots/qrange_T8_ROADS_fixed_mbr_USA_n10000_vary-c_comps_batch.pdf}
&\hspace{-1ex}\includegraphics[width=0.485\columnwidth]{../plots/qrange_T4_EDGES_fixed_mbr_USA_n10000_vary-c_comps_batch.pdf}\\
{\scriptsize query relative extent [\%]} &{\scriptsize query relative extent [\%]}\\
\end{tabular}
\caption{Batch query processing (window queries)}
\label{fig:vary-c_batch}
\end{figure}

\begin{figure}
\centering
\begin{tabular}{cc}
ROADS &EDGES\\
\hspace{-1ex}\includegraphics[width=0.485\columnwidth]{../plots/qrange_T8_ROADS_fixed_mbr_USA_n10000_p20000_batch_parallel.pdf}
&\hspace{-1ex}\includegraphics[width=0.485\columnwidth]{../plots/qrange_T4_EDGES_fixed_mbr_USA_n10000_p20000_batch_parallel.pdf}\\
{\scriptsize \# threads} &{\scriptsize \# threads}
\end{tabular}
\caption{Batch query parallel processing (window queries)}
\label{fig:vary-c_batch_parallel}
\end{figure}
}

\eat{
\subsection{Comparison with GeoSpark}
\label{sec:geospark}
\panos{In my opinion, this experiment does not fit the general narrative. I would remove it completely.}
Finally, we compare our proposed \twolevel grid index with
GeoSpark \cite{YuZS19}, one of the best-performing
distributed spatial data management systems according to
\cite{PandeyKNK18}.
Our goal is to show that, for the scale of benchmarking data \cite{EldawyM15}
that we and recent
papers \cite{PandeyKNK18,Li0ZY020,QiLJK20} use, 
in-memory indexing in a multi-core processing
machine is superior to using a system designed for cluster computing.
As our implementation is designed to run on a single machine,
we run GeoSpark in client mode, meaning that the driver and
Spark applications are both on the same machine. 
In addition, we used R-tree indexing in GeoSpark,
which performs best in query evaluation. 
We compared GeoSpark with \twolevel that uses a grid granularity of
1000x1000 and tested both single and multi-threaded versions for range queries. The experiments were conducted using the ROADS dataset on a machine with 64 GBs of RAM and a Intel(R) Core i7-4930K CPU clocked at 3.40GHz.%
\footnote{We do not own the platform where we ran the previous experiments, so we could not install GeoSpark on that machine.}
For each method, we average the cost of 100 (end-to-end) window queries, where the area of each query is 0.1\% of the area of the map.
Figure \ref{fig:geospark} shows that \twolevel always outperforms GeoSpark in terms of query performance by at least three orders of magnitude.
These results are consistent with the findings of 
\cite{PandeyKNK18}, where distributed spatial data management systems are shown to have a throughput of at most several hundred range queries per minute on data of similar scale.
In order to compare the two methods in a multi-threaded scenario on equal terms,
our approach evaluates the queries independently (i.e., not in batch). 
We observe the same trend as the number of cores increases. 

\begin{figure}[t]
\centering
  \includegraphics[width=0.55\columnwidth]{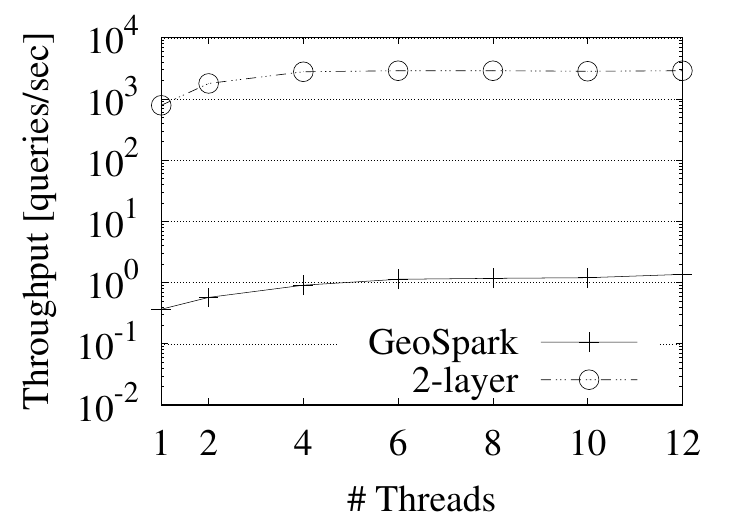}
  \caption{Window query performance comparison}
  \label{fig:geospark}
\end{figure}
}

\subsection{Updates Performance}
\ext{We also demonstrate the superiority of our two-layer partitioning
in updates. For this purpose,} we conducted an experiment using the real
datasets,
where we first
constructed the index by loading 90\% of the data in batch and then
measuring the cost of incrementally inserting the last 10\% of the data.
Table \ref{tab:updates_cost} compares the total update costs of the
competitor indices.
\rtree is two orders of magnitude slower than
the baseline \onelevel index and the cost of updates on \twolevel is
only a bit higher compared to the update cost on \onelevel.
Updates on \qtree are also slower compared to  \onelevel and
\twolevel, due to the tree traversal.

\eat{
  \begin{table}
\centering
\caption{Total update cost (sec)}
\label{tab:updates_cost}
\begin{tabular}{|c|c|c|c|}
\hline
\textbf{dataset}		&\textsf{R-tree}	&\onelevel		&\twolevel\\\hline\hline
ROADS		&5.34		&0.059	&0.068\\
EDGES					&19.8							&0.220	&0.241538\\
\hline
\end{tabular}
\end{table}
}

\begin{table}[t]
\centering
\caption{Total update cost (sec)\eat{ \todo{update me}}}
\label{tab:updates_cost}
\footnotesize
\begin{tabular}{|c|c|c|c|c|}
\hline
\textbf{dataset}		&\rtree					&\qtree						&\onelevel		&\twolevel\\\hline\hline
ROADS					&5.34					&0.76				 		&0.059				&0.068\\
EDGES					&19.8					&2.89		 				&0.267				&0.382\\
TIGER					&33.91					&4.63						&0.459				&0.634\\
\hline
\end{tabular}
\end{table}

\ext{
\subsection{Spatial Join Performance}
\label{sec:exps:joins}

\subsubsection{Two-layer Partitioning Join}
\label{sec:exps_join:mini-joins}
We first study the impact of our second layer of partitioning to the join computation. As discussed in Section~\ref{sec:join:mini-joins}, we assume that both input datasets are already indexed and that identical grids exist as the first layer of partitioning. We implemented the \emph{mj, base} method on top of our \twolevel scheme, as our basic mini-joins solution which adopts the mini-joins breakdown from Section~\ref{sec:join:mini-joins:breakdown} and the plane-sweep join approach from Section~\ref{sec:join:mini-joins:optimizations}. To demonstrate the effect of the comparison-saving optimizations from Section~\ref{sec:join:mini-joins:optimizations}, we also developed the \emph{mj, sans unecessary} and \emph{mj, sans redundant} variants, which extend \emph{mj, base}. Last, we implemented a mini-joins solution with all optimizations activated, denoted by \emph{mj, all opts} and \onelevel, a PBSM baseline solution that solely employs the first layer of partitioning.

Figure~\ref{fig:mini-joins} reports the breakdown of the total execution time while varying the number of partitions per dimension for the ROADS $\bowtie$ \revisions{ZCTA5} and EDGES $\bowtie$ \revisions{ZCTA5} queries. Note that we excluded the offline partitioning time but included the sorting time needed for adopting plane-sweep and saving on redundant comparisons.\footnote{Note that \onelevel employs the plane-sweep approach in \cite{BrinkhoffKS93} for partition-to-partition joins, similar to our \twolevel based methods.}
The results clearly show the benefit of employing our second layer of partitioning and the mini-joins breakdown compared to \onelevel. We also observe that employing all comparison-saving optimizations can further accelerate the join computation; the \emph{mj, all opts} always outperforms the rest tested methods.

\eat{
Figure~\ref{fig:mini-joins}, compare on sorting + sorting time the following:
\begin{itemize}
\item \onelevel
\item \twolevel base, i.e., mini-joins breakdown from Section~\ref{sec:join:mini-joins:breakdown} with plane-sweep discussed in the first paragraph of Section~\ref{sec:join:mini-joins:optimizations}
\item \twolevel base + plus the ``avoid unnecessary comparisons'' from Section~\ref{sec:join:mini-joins:optimizations}
\item \twolevel base + plus the ``avoid redundant comparisons'' from Section~\ref{sec:join:mini-joins:optimizations}
\item all opts
\end{itemize}
}
\begin{figure*}[t]
\centering
\begin{tabular}{c}
ROADS $\bowtie$ \revisions{ZCTA5}\\	
\hspace{-2ex}\includegraphics[width=0.9\linewidth]{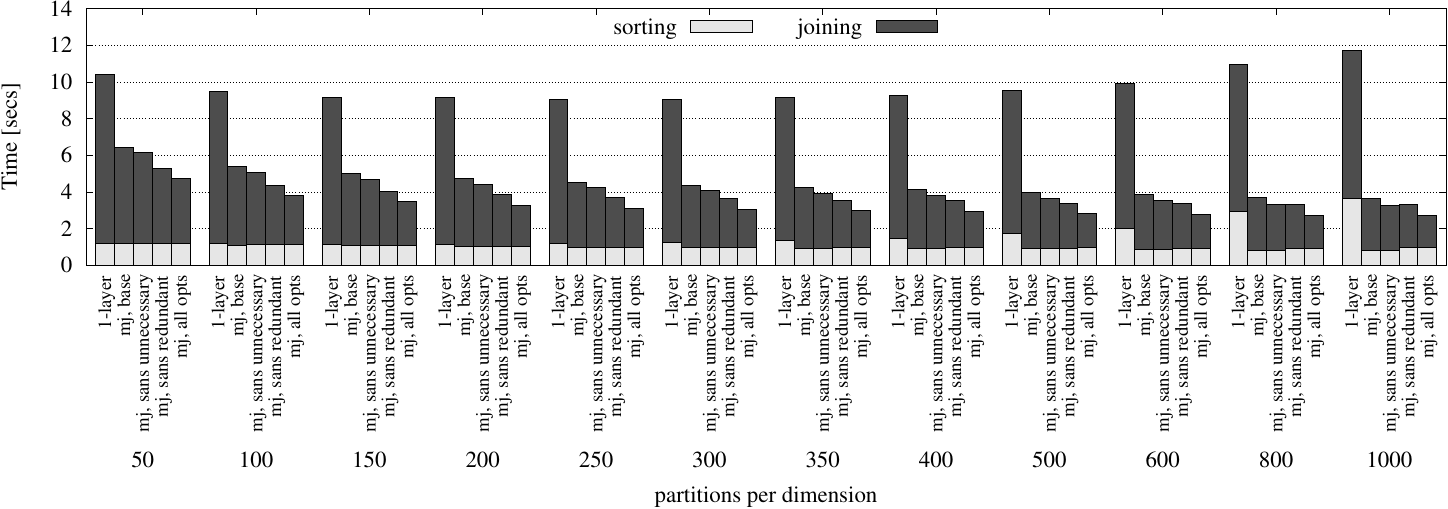}\\\\
EDGES $\bowtie$ \revisions{ZCTA5}\\
\hspace{-2ex}\includegraphics[width=0.9\linewidth]{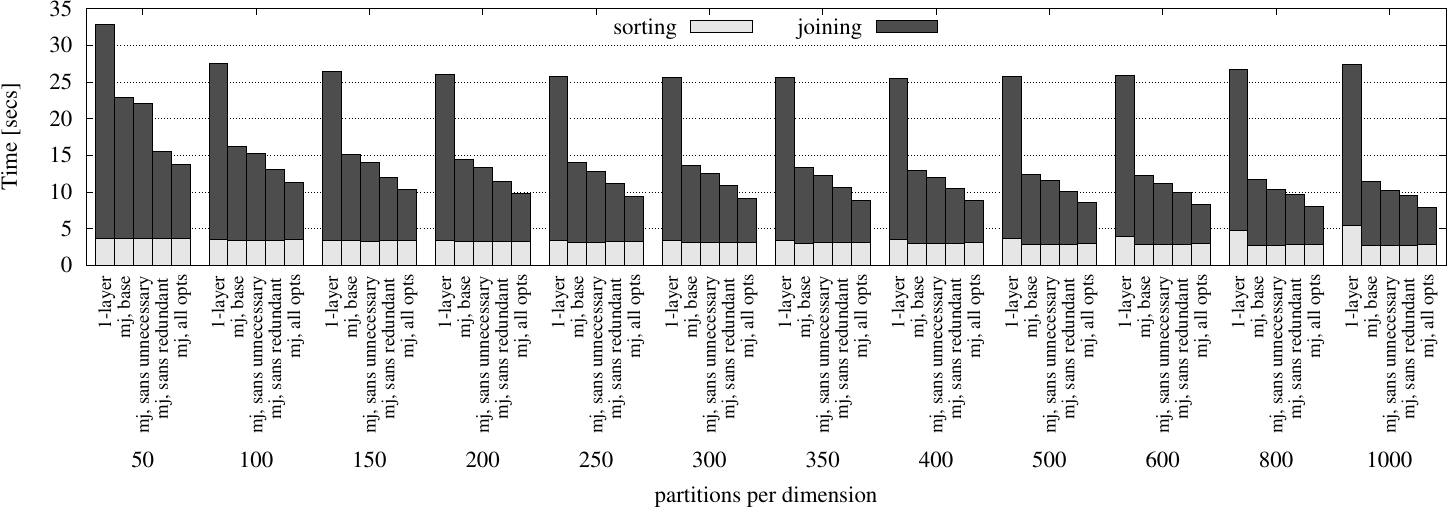}
\end{tabular}
\caption{Two-layer partitioning join on real datasets: mini-joins breakdown and optimizations.}
\label{fig:mini-joins}
\end{figure*}

\subsubsection{Both Inputs Indexed}
We then experiment with the three join strategies discussed in Section~\ref{sec:join:strategies}, starting with the setting where both input datasets are indexed by a \twolevel scheme. The granularity of each first layer grid is set according to our analysis in Section~\ref{sec:exps:tuning} \revisions{as powers of 2 (see Table~\ref{tab:granularities})}\eat{; i.e., the pre-existing \twolevel schemes are optimized for range queries}. We consider again the ROADS $\bowtie$ \revisions{ZCTA5} and EDGES $\bowtie$ \revisions{ZCTA5} queries. 

Figure~12\eat{\ref{fig:both-indexed}} reports the time breakdown of the tested approaches; as a baseline, we also included a traditional R-tree join method \cite{BrinkhoffKS93}. We tested the straightforward approach of re-indexing one of the inputs using a temporary \twolevel so that both indices employ identical grids and being able to use the \emph{mj, all opts} join method from the previous section. The figure shows the results for re-indexing ROADS and EDGES, while re-indexing \revisions{ZCTA5} is omitted. As \revisions{ZCTA5} contains significantly larger rectangles than ROADS and EDGES (see Table~\ref{tab:real}), indexing these rectangles under the very fine grids used by \twolevel for ROADS and EDGES incurs a high replication ratio and hence, high partitioning costs (over 300 secs). The figure also includes the two variants for the online transformation of the \twolevel scheme in ROADS and EDGES to the grid of \revisions{ZCTA5}, proposed in Section~\ref{sec:join:strategies:bothindexed}. The results clearly show the benefit the online transformation compared to re-indexing one of the inputs and in particular, the variant when the new \twolevel scheme is not materialized as such an approach completely eliminates the partitioning costs. 
\eat{
\begin{itemize}
\item re-index finer to coarse
\item re-index coarse to finer
\item transformation, materialized
\item transformation, non-materialized
\item \rtree join
\end{itemize}
}
\eat{
\begin{table*}
\centering
\scriptsize
\caption{\todo{update}}
\hspace*{-8ex}
\begin{tabular}{|c|c|c|c|c|c|}
\hline
\multirow{3}{*}{\textbf{query $R\bowtie S$}}	&\multicolumn{5}{c|}{\textbf{approaches}}\\\cline{2-6}
											&\multirow{2}{*}{\rtree join}		&\multicolumn{2}{c|}{\textbf{direct re-indexing}}		&\multicolumn{2}{c|}{\textbf{online transformation}}\\\cline{3-6}
											&								&$R$									&$S$ 						&materialized			&non-materialized\\\hline\hline
ROADS $\bowtie$ \revisions{ZCTA5}						&								&0.702958+0.959067+1.94067									&254.87+2.12302+1.87445							&0.3674164444+0.9141344444+1.61855						&1.144544444+1.659923333\\
EDGES $\bowtie$ \revisions{ZCTA5}						&								&2.04295+3.02143+5.63945			&761.49+5.73909+5.48015	&1.44424+2.802431111+4.757371111	&2.412157778+4.634305556\\\hline
\end{tabular}
\end{table*}
}
\eat{
\begin{figure}[t]
\begin{tabular}{cc}
\multicolumn{2}{c}{\includegraphics[width=0.8\columnwidth]{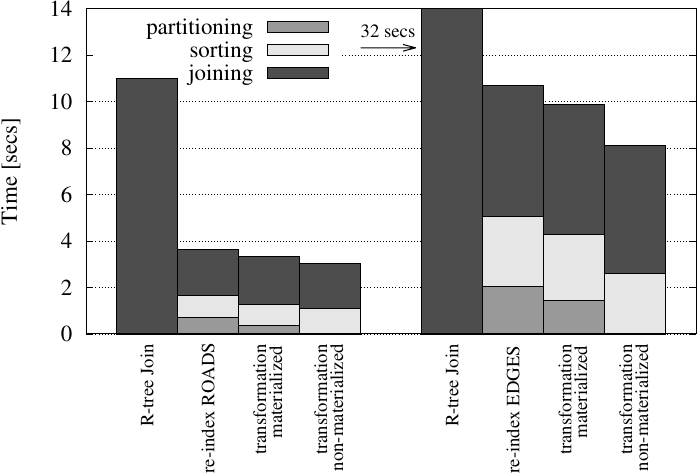}}\\
~~~~~~~~~~~~~ROADS $\bowtie$ \revisions{ZCTA5} &EDGES $\bowtie$ \revisions{ZCTA5}\\
\end{tabular}
\caption{`Both Inputs Indexed' setting on real datasets\todo{update \rtree join}. Re-index \revisions{ZCTA5} approach omitted due its high online partitioning costs ($> 300$ secs).}
\label{fig:both-indexed}
\end{figure}
}

\subsubsection{One Input Indexed}
We consider once again the ROADS $\bowtie$ \revisions{ZCTA5} and EDGES $\bowtie$ \revisions{ZCTA5} queries when only one input is indexed by our \twolevel scheme. 
We tested three solutions in this context. The first is a classic index-based join; we denote this approach as \emph{for-loop, probe}.
The second approach \emph{grid $10\times 10$, probe} extends this idea by partitioning the unindexed input with a coarse grid (we used a $10\times 10$ one) and then executing the window range queries per grid cell. Finally, we also employed the `Both Inputs Indexed' approach by indexing the unindexed input online.

Figure~13\eat{\ref{fig:one-indexed}} reports the breakdown of the execution time; we distinguish between two cases for each query, depending on which input is already indexed. Note that we omit the `Both Inputs Indexed' approach when \revisions{ZCTA5} has to be indexed, similarly to previous section. We discuss the ROADS $\bowtie$ \revisions{ZCTA5} query as the findings are the same for EDGES $\bowtie$ \revisions{ZCTA5}. When \revisions{ZCTA5} is indexed, we observe that building online a temporary \twolevel scheme on ROADS with an identical grid to \revisions{ZCTA5} is in fact the best solution. This is mainly because \emph{mj, all opts} used to compute the join drastically reduces the joining time. In contrast, when ROADS is pre-indexed, the best approach is \emph{grid $10\times 10$, probe}\eat{ to examine the rectangles in \revisions{ZCTA5} divided by a grid} which improves the cache locality when probing the index on ROADS. As \revisions{ZCTA5} contains significantly fewer rectangles than ROAD, the cost of constructing such a grid is negligible compared to applying the same idea on ROADS when \revisions{ZCTA5} is pre-indexed. To sum up, when one of the inputs is pre-indexed by our \twolevel scheme, an $R \bowtie S$ spatial join can be efficiently computed using the `Both Inputs Indexed' strategy if the cost of indexing the second input is expected to be low (due to containing small rectangles or using a coarse first layer grid) or the grid-based probe approach otherwise.
\eat{
\begin{itemize}
\item simple for-loop based probes
\item grid-based probes
\item batch processing powered (?)
\item both-indexed approach
\end{itemize}
}
\eat{
\begin{figure}[t]
\begin{tabular}{cccc}
\multicolumn{2}{c}{ROADS $\bowtie$ \revisions{ZCTA5}} &\multicolumn{2}{c}{EDGES $\bowtie$ \revisions{ZCTA5}}\\
\multicolumn{2}{c}{\includegraphics[width=0.7\columnwidth]{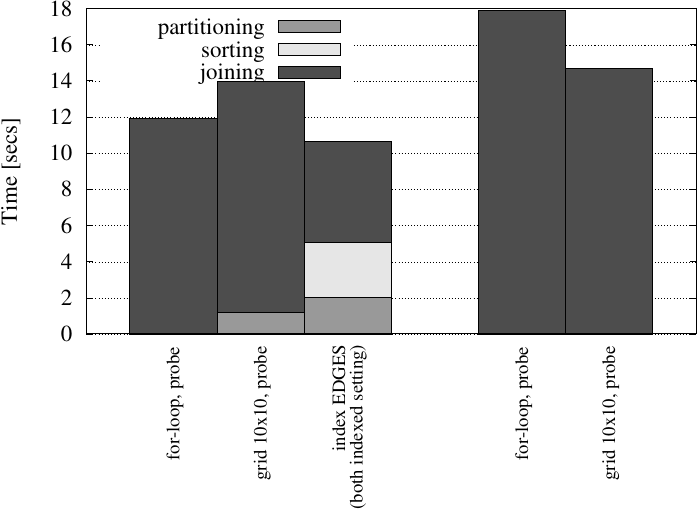}}\\
&\multicolumn{2}{c}{\includegraphics[width=0.7\columnwidth]{plots_new/setting_one-indexed_T10-T4_XOR.pdf}}\\
~~~~~~~~~~~~~\revisions{ZCTA5} indexed &EDGES indexed\\
\end{tabular}
\caption{`One Input Indexed' setting on real datasets}
\label{fig:one-indexed}
\end{figure}
}

\begin{figure*}[t]
\begin{minipage}[b]{.35\textwidth}
\begin{tabular}{cc}
\multicolumn{2}{c}{\includegraphics[width=0.83\columnwidth]{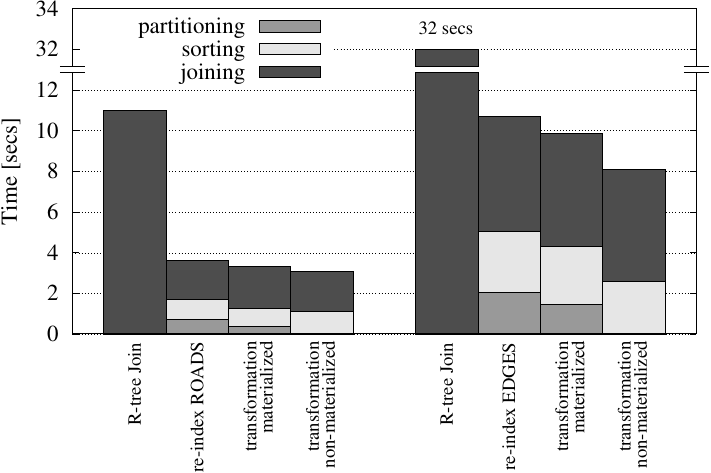}}\\
\hspace*{5ex}{\footnotesize ROADS $\bowtie$ \revisions{ZCTA5}} &{\footnotesize EDGES $\bowtie$ \revisions{ZCTA5}}\\
\end{tabular}
\label{fig:both-indexed}
\captionof{figure}{`Both Inputs Indexed' setting on real datasets; re-indexing \revisions{ZCTA5} omitted due to high online partitioning costs\eat{ ($> 300$ secs)}.}
\end{minipage}
\hspace{2ex}
\begin{minipage}[b]{.6\textwidth}
\begin{tabular}{cc}
ROADS $\bowtie$ \revisions{ZCTA5} &EDGES $\bowtie$ \revisions{ZCTA5}\\
\includegraphics[width=0.49\columnwidth]{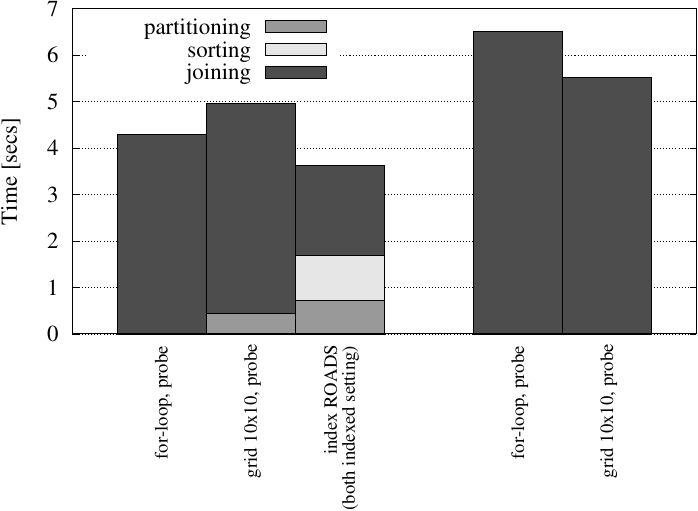}
&\hspace*{-4ex}\includegraphics[width=0.49\columnwidth]{plots_new/setting_one-indexed_T10-T4_XOR.pdf}\\
~~~~~~~~~{\footnotesize \revisions{ZCTA5} indexed}~~~~~{\footnotesize ROADS indexed}
&~~~~~{\footnotesize \revisions{ZCTA5} indexed}~~~~~{\footnotesize EDGES indexed}\\
\end{tabular}
\label{fig:one-indexed}
\captionof{figure}{`One Input Indexed' setting on real datasets; indexing \revisions{ZCTA5} omitted due to high online partitioning costs.}
\vspace{0.5ex}
\end{minipage}
\end{figure*}

\subsubsection{No Input Indexed}
Lastly, we consider the setting when none of the input\eat{ dataset}s is indexed. As discussed in Section~\ref{sec:join:strategies:noindex}, we directly construct in this case, two temporary \twolevel schemes under an identical first layer grid and then use \emph{mj, all opts} to compute the join.  Figure~\ref{fig:no-index} shows the effect of further optimizing this approach with the space reduction optimization proposed at the end of Section~\ref{sec:join:strategies:noindex}, while varying 
the grid granularity.
We denote this method as \emph{mj, all opts + s-opt}. We observe that the space optimization enhances the join computation, especially when a very fine \eat{first layer }grid is used.
\eat{
\begin{itemize}
\item all opts from Section~\ref{sec:exps_join:mini-joins}
\item all opts from Section~\ref{sec:exps_join:mini-joins} + space reduction (formally known as decomposition)
\end{itemize}
}
\begin{figure}[t]
\centering
\begin{tabular}{c}
ROADS $\bowtie$ \revisions{ZCTA5}\\	
\hspace{-2ex}\includegraphics[width=\columnwidth]{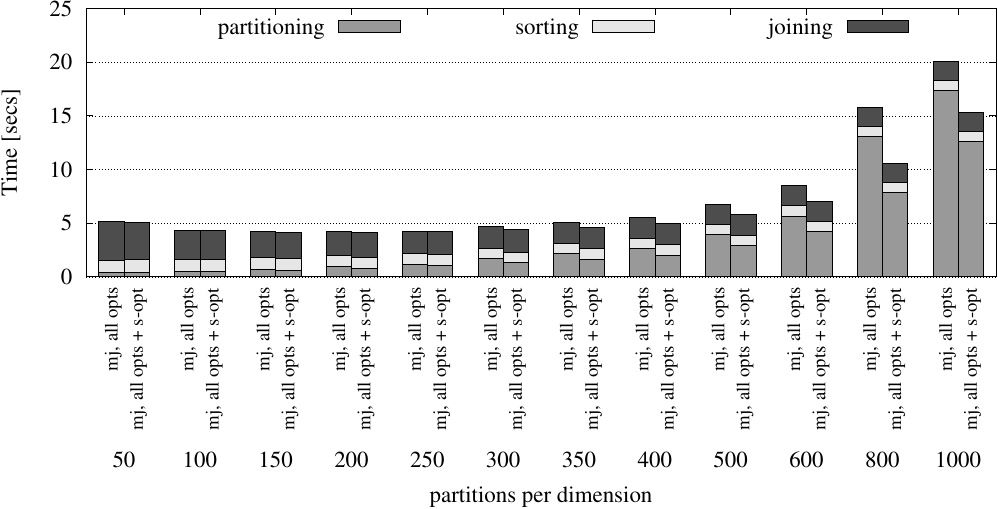}\\\\
EDGES $\bowtie$ \revisions{ZCTA5}\\
\hspace{-2ex}\includegraphics[width=\columnwidth]{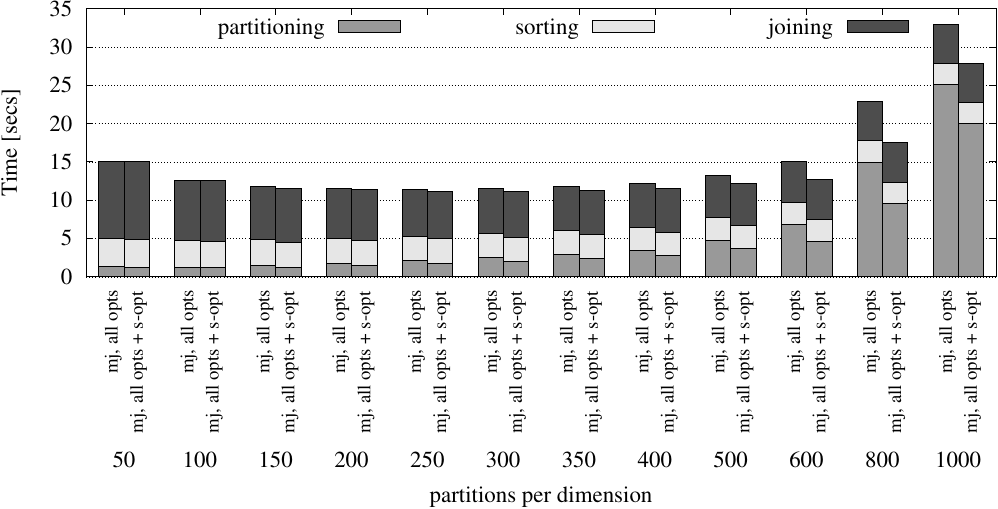}
\end{tabular}
\caption{`No Input Indexed' setting on real datasets}
\label{fig:no-index}
\end{figure}

\eat{

\subsection{Join Queries Performance}
\todo{New experiments for joins:
\begin{itemize}
\item Effectiveness of optimizations for the case of both input indexed by a two-layer partitioning
\item Compare different join strategies - R-tree join?
\item Dittrich is out of the picture
\end{itemize}
}

\subsubsection{No-Index}
\begin{figure*}[t]
\eat{
\begin{center}
\begin{small}
\fbox{\parbox{480pt}
{
\hspace{2ex}
{\footnotesize HINT for \overlaps}
\includegraphics[width=0.06\columnwidth]{../../figures/blue_0_2.pdf}
\hspace{2ex}
{\footnotesize HINT one setup for all}
\includegraphics[width=0.06\columnwidth]{../../figures/blue_0_1.pdf}
\hspace{2ex}
{\footnotesize HINT$^m$ for \overlaps}
\includegraphics[width=0.06\columnwidth]{../../figures/brown_8.pdf}
\hspace{2ex}
{\footnotesize HINT$^m$ one setup for all}
\includegraphics[width=0.06\columnwidth]{../../figures/brown_8_0_4.pdf}
}
}
\end{small}
\end{center}
}
\begin{tabular}{c}
ROADS $\bowtie$ \revisions{ZCTA5}\\	
\hspace{-2ex}\includegraphics[width=\linewidth]{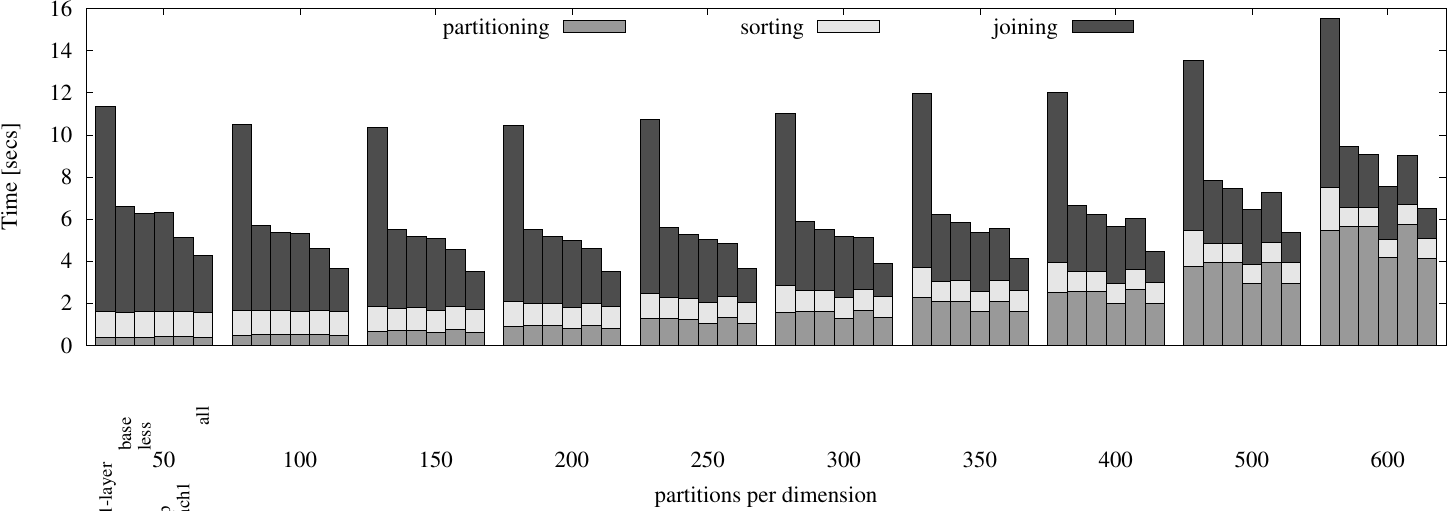}\\\\
EDGES $\bowtie$ \revisions{ZCTA5}\\
\hspace{-2ex}\includegraphics[width=\linewidth]{plots_new/T10-T4_noindex.pdf}
\end{tabular}
\caption{\todo{No-index strategy: Optimizations' effect}}
\label{fig:opts}
\end{figure*}

\subsubsection{One Index}
\eat{
\begin{table}
\centering
\caption{\todo{update}}
\begin{tabular}{|c|c|c|}
\hline
\textbf{method}	&\revisions{ZCTA5} $\bowtie$ ROADS		&ZCODE $\bowtie$ EDGES\\\hline\hline
base				&6.49172						&17.8602\\
grid10			&5.5016						&\textbf{14.6904}\\
grid100			&\textbf{5.46597}			&15.0675\\
grid1000			&5.46608						&15.0697\\\hline
\end{tabular}
\end{table}
}
\begin{table}
\centering
\caption{\todo{update}}
\begin{tabular}{|c|c|c|c|c|}
\hline
\textbf{query}			&base		&grid10				&grid100				&grid1000\\\hline\hline
\revisions{ZCTA5} $\bowtie$ ROADS	&6.49		&5.50				&\textbf{5.46}		&5.47\\
ROADS $\bowtie$ \revisions{ZCTA5}	&&&&\\
ZCODE $\bowtie$ EDGES	&17.9		&\textbf{14.7}		&15.1				&15.1\\
EDGES $\bowtie$ \revisions{ZCTA5}	&\textbf{11.5}		&13.0				&12.9				&13.1\\\hline
\end{tabular}
\end{table}
}
}

\section{Conclusions}
\label{sec:conclusion}
We presented a secondary partitioning approach that can
be applied to SOP indices, such as grids, and
divides the MBRs within each spatial partition to
four classes.
Our approach reduces the number of comparisons during
range query evaluation and avoids the generation (and elimination) of
duplicate results.
  We show how our approach can also be used for duplicate result avoidance
  in spatial intersection joins that are evaluated using the state-of-the-art
  PBSM algorithm.
For both range queries and joins, we show how redundant
computations can also be avoided.
Our experimental findings confirm the superiority of our
approach compared to the state-of-the-art duplicate result elimination
method \cite{DittrichS00}.
We also show that a grid equipped with our method outperforms
other indices (such as the \qtree and \rtree) by up to one order of
magnitude\eat{ and showed its scalability to multiple query
evaluation in parallel}.
\ext{The cost of spatial intersection joins is also reduced by a
  significant factor (about 50\%) with the help of our secondary
  partitioning technique and the use of our optimized
  partition-to-partition join algorithms.  
}
Directions for future work include the evaluation of 
distance-based queries, e.g., $k$-NN or distance joins,
the application of our approach to
distributed spatial data management systems \revisions{and
to indexing trajectories for retrieval and join problems.}



%

\eat{
\ifCLASSOPTIONcompsoc
  \section*{Acknowledgments}
\else
  \section*{Acknowledgment}
\fi

D. Tsitsigkos and M. Terrovitis were supported by EU’s Horizon 2020 programme, MORE project (Grant Agreement No. 957345).
P. Bouros was at the time, a Carl-Zeiss Stiftungsprofessor for “Big Data: In-Memory Databases and Data Analytics”.
N. Mamoulis was supported by the Hellenic Foundation for Research and Innovation (HFRI) under the ``2nd Call for HFRI Research Projects to support Faculty Members \& Researcher'' (Project No. 2757). 
The authors gratefully acknowledge the computing time on the supercomputer Mogon at Johannes Gutenberg University Mainz (hpc.uni-mainz.de).
}



\bibliographystyle{IEEEtran}
\bibliography{sjoin}
%
\eat{

}

%
\eat{
\vspace*{-7ex}
\begin{IEEEbiography}[{\includegraphics[width=1in,height=1.25in,clip,keepaspectratio]{photos/tsitsi}}]{Dimitrios Tsitsigkos}
received his bachelor and master degree from the National and Kapodistrian University of Athens in 2012 and 2016, respectively. Currently, He is a software engineer at the Institute for the management of Information Systems (IMSI) of RC “Athena” and a PhD candidate at the University of Ioannina, in Greece. His research focuses on join operators for complex data.
\end{IEEEbiography}

\vspace*{-7ex}
\begin{IEEEbiography}[{\includegraphics[width=1in,height=1.25in,clip,keepaspectratio]{photos/panos}}]{Panagiotis Bouros}
received his diploma and doctorate degree from the School of Electrical and Computer Engineering at the National Technical University of Athens, Greece, in 2003 and 2011, respectively. Currently, he is an assistant professor at the Institute of Computer Science, Johannes University Gutenberg Mainz, Germany. Before, he held 
positions in Denmark, Germany and Hong Kong. 
His research focuses on managing \eat{and querying }complex data types and
query processing.
\end{IEEEbiography}

\vspace*{-7ex}
\begin{IEEEbiography}[{\includegraphics[width=1in,height=1.25in,clip,keepaspectratio]{photos/ntinos}}]{Konstantinos Lampropoulos}
received his Computer Science and Engineering diploma from the University of Ioannna in 2019. He is currently a PhD candidate at the same institution\eat{ under the supervision of Professor Nikos Mamoulis}. His research interests include main-memory database systems and query processing on big data.
\end{IEEEbiography}

\vspace*{-7ex}
\begin{IEEEbiography}[{\includegraphics[width=1in,height=1.25in,clip,keepaspectratio]{photos/nikos}}]{Nikos Mamoulis}
received his diploma in computer engineering and informatics in 1995
from the University of Patras, Greece, and his PhD in computer science
in 2000 from HKUST. He is currently a faculty member at the University
of Ioannina, Greece. Before, he was a professor at the Department of
Computer Science, University of Hong Kong. His research focuses on
managing and mining of complex data types.
\end{IEEEbiography}

\vspace*{-7ex}
\begin{IEEEbiography}[{\includegraphics[width=1in,height=1.25in,clip,keepaspectratio]{photos/mterCut}}]{Manolis Terrovitis}
is a principal researcher at the Institute for the Management of Information Systems (IMSI) of the ``Athena'' Research and Innovation Centre in Information, Communication and Knowledge Technologies in Greece. He received his PhD in 2007 from the National Technical University of Athens. His main research interests lie in the areas of data privacy, indexing and query evaluation.
\end{IEEEbiography}
}



\end{document}